%% file: lmc_long.tex
\documentstyle[graphicx]{mn}

\newcommand\aproxgt{\mathrel{%
      \rlap{\raise 0.511ex \hbox{$>$}}{\lower 0.511ex \hbox{$\sim$}}}}
\newcommand\aproxlt{\mathrel{%
      \rlap{\raise 0.511ex \hbox{$<$}}{\lower 0.511ex \hbox{$\sim$}}}}
\def\errtwo#1#2#3{${#1}^{+#2}_{-#3}$}

\newcommand\xone{LMC~X-1}
\newcommand\xthree{LMC~X-3}
\newcommand\asca{\textsl{ASCA}}
\newcommand\rxte{\textsl{RXTE}}
\newcommand\pca{\textsl{PCA}}
\newcommand\hexte{\textsl{HEXTE}}
\newcommand\asm{\textsl{ASM}}
\newcommand\sis{\textsl{SIS}}
\newcommand\gis{\textsl{GIS}}
\newcommand\Msun{{\rm M}_\odot}

\begin{document}

\title{A Good Long Look at the Black Hole Candidates \xone\ and \xthree}
\author[M. A. Nowak et al.]{
  M. A. Nowak$^1$, J. Wilms$^2$, W. A. Heindl$^3$, K. Pottschmidt$^2$, 
  J. B. Dove$^{4,5}$, \newauthor M. C. Begelman$^{1,6}$ \\
$^1$ JILA, University of Colorado, Campus Box 440, Boulder, CO
  80309-0440, U.S.A.; mnowak@rocinante.colorado.edu \\
$^2$ Institut f\"ur Astronomie und Astrophysik,
  Abt.~Astronomie, Waldh\"auser Str. 64, D-72076 T\"ubingen, Germany; \\
   wilms@astro.uni-tuebingen.de, katja@astro.uni-tuebingen.de \\
$^3$ Center for Astronomy and Space Sciences, University of
  California at San Diego, La Jolla, CA 92093, U.S.A.; \\
  wheindl@mamacass.ucsd.edu  \\
$^4$ Center for Astronomy and Space Astrophysics, University of
  Colorado, Boulder, CO 80309-389, U.S.A.; dove@casa.colorado.edu \\
$^5$ also, Dept. of Physics, Metropolitan State College of
  Denver, C.B. 69, P.O. Box 173362, Denver, CO 80217-3362, U.S.A. \\
$^6$ also, Dept. of Astrophysics and Planetary Sciences,
  University of Colorado, Boulder 80309, U.S.A.; mitch@jila.colorado.edu }
\date{Accepted 2000 ; Received 2000; in original form}

\maketitle

\label{firstpage}

\begin{abstract}  
  \xone\ and \xthree\ are the only known persistent stellar-mass black hole
  candidates that have almost always shown spectra that are dominated by a
  soft, thermal component.  We present here results from 170\,ksec long
  Rossi \mbox{X-ray} Timing Explorer (\rxte) observations of these objects,
  taken in 1996~December, where their spectra can be described by a disc
  black body plus an additional soft ($\Gamma\sim 2.8$) high-energy
  power-law (detected up to energies of 50\,keV in \xthree).  These
  observations, as well as archival Advanced Satellite for Cosmology and
  Astrophysics (\asca) observations, constrain any narrow Fe line present
  in the spectra to have an equivalent width $\aproxlt 90$\,eV.  Stronger,
  broad lines ($\approx 150$\,eV EW, $\sigma \approx 1$\,keV) are
  permitted.  We also study the variability of \xone.  Its \mbox{X-ray}
  power spectral density (PSD) is approximately $\propto f^{-1}$ between
  $10^{-3}$ and $0.3$\,Hz with a root mean square (rms) variability of
  $\approx 7\%$.  At energies $>5$\,keV the PSD shows evidence of a break
  at $f > 0.2$\,Hz, possibly indicating an outer disc radius of $\aproxlt
  1000~GM/c^2$ in this likely wind-fed system.  Furthermore, the coherence
  function $\gamma^2(f)$, a measure of the degree of linear correlation,
  between variability in the $> 5$\,keV band and variablity in the lower
  energy bands is extremely low ($\aproxlt 50\%$). We discuss the
  implications of these observations for the mechanisms that might be
  producing the soft and hard X-rays in these systems.
\end{abstract}


\section{Introduction}\label{sec:intro}

Since the discovery of Cygnus~X-1 in 1964 \cite{bowyer:65a}, the study of
galactic black hole candidates (BHCs) has shown that these objects display
a large variety of states which are characterized by their distinct
spectral shapes and temporal behaviours. The most important states which
have been identified are the ``low/hard state'', which is characterized by
a hard X-ray spectrum with a photon index $\Gamma=1.7$ and large root mean
square (rms) variability $\aproxgt 30\%$ (Tanaka \& Lewin 1995; Nowak
1995\nocite{tanaka:95a,nowak:95a}; and references therein), and the
``high/soft state'', which is spectrally softer ($\Gamma \sim 2.5$) and
exhibits less variability. The fractional Eddington luminosity of sources
in the soft state tends to be higher than that for sources in the hard state
(Nowak 1995\nocite{nowak:95a}, and references therein).

The soft state has been observed in steady sources such as \xone\ and
\xthree\ 
\cite{treves:88a,ebisawa:89a,treves:90a,ebisawa:93a,schmidtke:99a}, in
recurring transients such as GX~339$-$4 \cite{grebenev:93a}, and in a
number of transients such as Nova Muscae \cite{miyamoto:94a}.  The
persistent BHC Cygnus~X-1 has been observed to switch between the hard and
the soft state (albeit with $kT\sim 0.3$\,keV for the soft state), with the
total X-ray luminosity staying roughly constant \cite{cui:97b,zhang:97b}.

A great deal of observational attention has been focused on the more
commonly observed hard state, since most of the brighter galactic BHCs only
occasionally transit to the soft state.  Only two of the persistent nearby
BHCs, \xone\ and \xthree, until recently have always been observed in the
soft state. Wilms et al. \shortcite{wilms:99b}, hereafter Paper II, present
evidence that \xthree\ periodically transits into the low/hard state.  In
this work we present 170\,ksec long Rossi \mbox{X-ray} Timing Explorer
(\rxte) observations of both \xone\ and \xthree\ during high/soft X-ray
states.

\xthree\ is a highly variable BHC with a probable 9\,${\rm M}_{\odot}$
compact object mass \cite{cowley:83a}.  Its luminosity has been observed to
be as high as $4\times 10^{38}\,{\rm erg~s^{-1}}$, which is $\approx 30\%$
of its Eddington luminosity, and its soft X-ray flux ($\sim 1$--$9$\,keV)
is variable by a factor of more than four on long time scales (see Paper
II).  \xthree\ also has exhibited a strong 99- or 198-day periodicity in
its soft X-ray flux \cite{cowley:91a,cowley:94a}, which is also apparent in
the \asm\ monitoring, albeit with a period that varies over time (Paper
II).  Previously we had suggested that this periodicity might be associated
with a warped, precessing accretion disc \cite{wilms:99a}; however, with
the recently observed ``state changes'' to a low/hard flux it now seems
likely that a systematic variation of the accretion rate is an important
part of this long-term variability.

\xone\ is also a good candidate for a black hole.  Using a large number of
\textsl{ROSAT} HRI observations, Cowley et al. \shortcite{cowley:95a} were
able to identify the counterpart with ``star number 32'' of Cowley et al.
\shortcite{cowley:78a}.  This object has a mass function of only
$f=0.144\,\Msun$, but including other evidence the mass of the compact
object appears to be $M>4\,\Msun$ \cite{hutchings:87a}, and probably
$\aproxgt 6\,\Msun$ \cite{cowley:95a}.  The luminosity of the object is
typically about $2\times 10^{38}\,{\rm erg~s^{-1}}$ \cite{long:81a}.  Although
small differences between \xone\ and \xthree\ are apparent, their spectra
and short-time temporal behaviours have historically been quite similar.
\xone, however, does not exhibit any obvious periodic behaviour in its long
term X-ray lightcurve.

We have monitored \xone\ and \xthree\ with \rxte\ in three to four weekly
intervals since the end of 1996 in order to enable a systematic study of
the soft state. The campaign started with 170\,ksec long observations of
both sources, and in this paper we present results from the spectral and
temporal analysis of these long observations.  In addition, we consider
spectral results of archival Advanced Satellite for Cosmology and
Astrophysics (\asca) observations.  Preliminary results of our analyses
have already appeared elsewhere \cite{wilms:99a,wilms:99c}.  Results from
the monitoring observations are presented in Paper II.

The remainder of this paper is structured as follows. We start with a
description of our \rxte\ data analysis methodology
(\S\ref{sec:rxte}). We then present the results from the spectral
analysis from \rxte\ (\S\ref{sec:spectra}) and \asca\ 
(\S\ref{sec:asca}). The \rxte\ timing analysis is discussed in
section~\ref{sec:timing}.  We discuss our results in the context of
current physical models for the soft state (\S\ref{sec:discuss}) and
then summarize the paper (\S\ref{sec:summary}).

\section{RXTE Data Analysis Methods}\label{sec:rxte}
Onboard \rxte\ are two pointed instruments, the Proportional Counter Array
(\pca) and the High Energy X-ray Timing Experiment (\hexte), as well as the
All Sky Monitor (\asm). We used the standard \rxte\ data analysis software,
ftools~4.2. Spectral modeling was performed mostly with XSPEC, version
10.00z \cite{arnaud:96a}. We used essentially the same data extraction and
analysis strategy as in our analyses of \rxte\ data from GX~339$-$4 and
V1408~Aql (4U1957$+$11); therefore, we only present a brief summary of this
strategy. For detailed information we refer to Wilms et al.
\shortcite{wilms:99aa} and Nowak \& Wilms \shortcite{nowak:99d}.

The \pca\ consists of five co-aligned Xenon (with an upper Propane layer)
proportional counter units (PCUs) with a total effective area of about
$6500\,\mbox{cm}^2$. The instrument is sensitive in the energy range from
2\,keV to $\sim 100$\,keV \cite{jahoda:96b}, although the response matrix
is best calibrated in the energy range of $\approx$2.5--30\,keV.  We chose
to increase the signal to noise ratio of our data by analyzing top Xenon
layer data only.  Background subtraction of the \pca\ data was performed
analogously to our previous study of Cyg~X-1 \cite{dove:98a}. To reduce the
uncertainty of the \pca\ background model, we ignored data measured in the
30\,minutes after South Atlantic Anomaly (SAA) passages. The XSPEC
`corrfile' facility was used to renormalize the background file for all
observations, typically by 1--2\%.

For short ($\aproxlt 10$\,ks) segments of observations, we assumed the
uncertainty of the data to be purely from a Poisson distribution.  For
analysis of the full ($>100$\,ks) observations, we accounted for the
remaining detector calibration uncertainty by using the energy dependent
systematic errors described by Wilms et al. \shortcite{wilms:99aa}.

\hexte\ consists of two clusters of four NaI/CsI-phoswich scintillation
counters that are sensitive from 15 to 250\,keV. A full description of the
instrument is given by Rothschild et al. \shortcite{rothschild:98a}.
Background subtraction is done by source-background switching. We used the
standard \hexte\ response matrices of 1999~August, and considered data
measured above 20\,keV.

\section{RXTE Spectral Analysis}\label{sec:spectra}

\subsection{\xthree}\label{sec:x3xte}

\input{Table1}

\begin{figure}
\centerline{\includegraphics[width=0.45\textwidth]{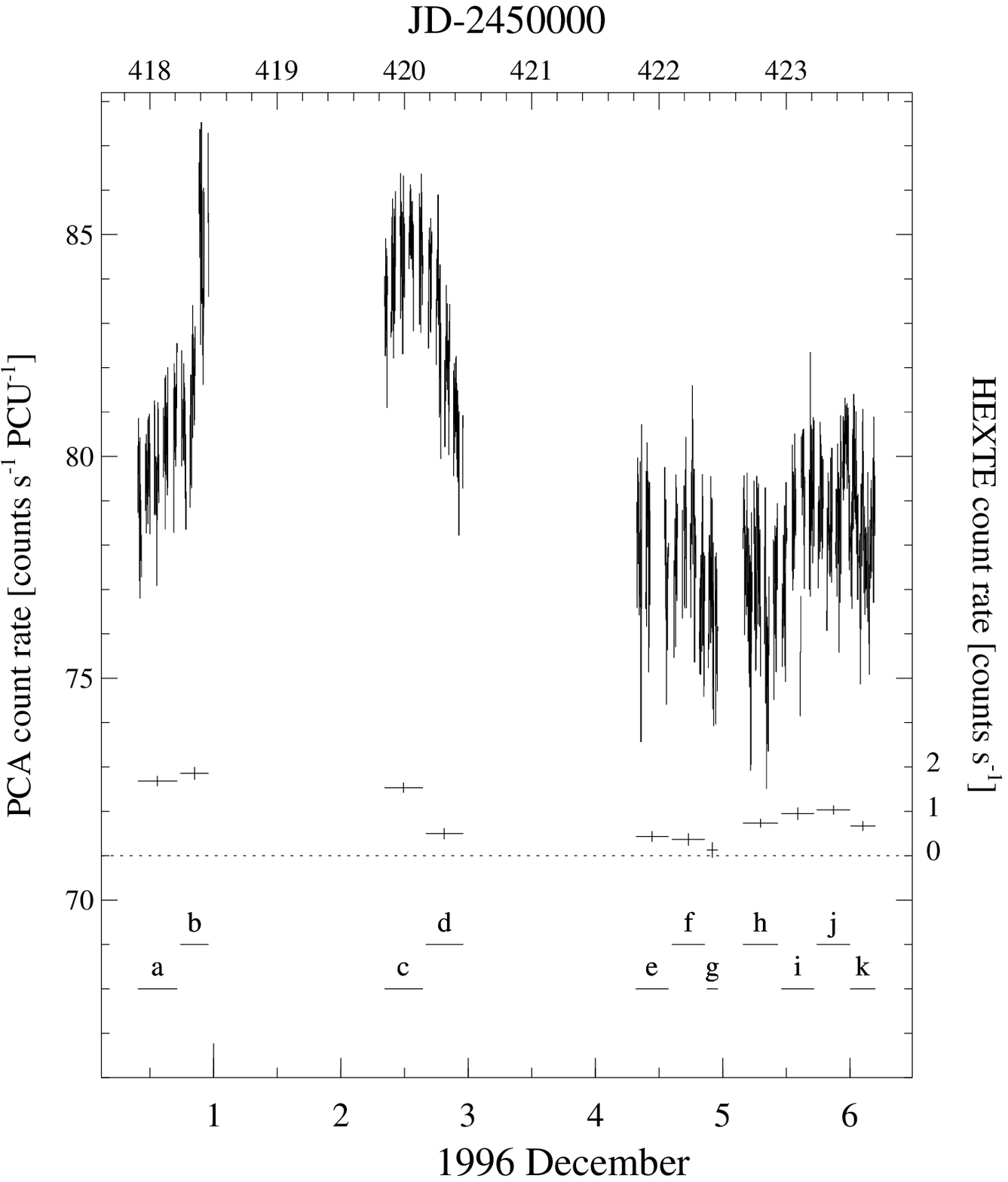}}
\caption{Background-subtracted \rxte\ \pca\ lightcurve of \xthree,
  rebinned to a resolution of 64\,s (left $y$-axis), and total
  background subtracted \hexte\ count rate (sum of \hexte\ clusters~A
  and~B; right $y$-axis).  The dashes indicate the individual
  observations of Table~\protect{\ref{tab:x3log}}.
  \label{fig:x3longlc}}
\end{figure}

\rxte\ observed \xthree\ during three observing blocks on
1996~November 30, 1996~December~2, and 1996~December~4 and~5.  A log
of the observations can be found in Table~\ref{tab:x3log}. During part
of the observations, some detectors were turned off; however, we added
all good data together and combined all background models. The total
response matrix was the average of the respective estimates for each
individual detector combination, weighted by the fraction of photons
measured during the time that such a combination was active.

\input{Table2}

The lightcurve of \xthree\ during the long observation is displayed in
Fig.~\ref{fig:x3longlc}. During the first part of the observation the
source is characterized by a slight increase of the \pca\ count rate from
$\sim 80$ to $\sim 85$\,cps per PCU.  The source was at a relatively steady
level of $\sim 77$\,cps per PCU during the last part of the observations.
This variation was also observed in the \hexte\ (Fig.~\ref{fig:x3longlc}).
To see whether this variation was also detectable spectroscopically, we
first analyzed separately the observations indicated in
Table~\ref{tab:x3log}.  For this analysis, we ignored data below \pca\ 
channel 7 and above \pca\ channel 50, which corresponds to an energy
bandpass from 3.6--20\,keV. (Here and throughout our spectral discussions,
PCA channel number refers to the `standard\_2f' channels, 1--129, as read
by XSPEC.)

\begin{figure}
\centerline{\includegraphics[width=0.5\textwidth]{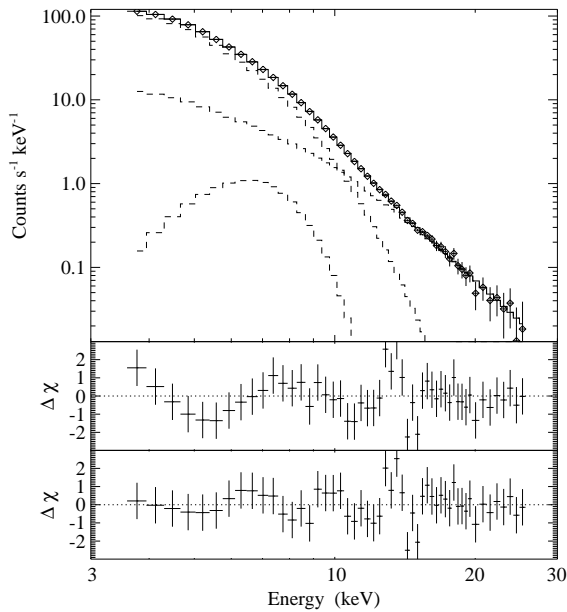}}
\caption{Top: Observed spectrum and disc black body plus
  power law plus broad iron line model fit to the long observation of
  \xthree. Dashed lines show the individual model components folded through
  the detector response.  Middle: Residuals in units of $\sigma$ between
  the data and the model, with a narrow iron line. Bottom: Residuals for
  the best fit model, including a very broad iron line.  (Residuals for the
  {\tt compps} model are virtually identical.) \label{fig:x3longspec}}
\end{figure}

We first modeled the spectrum via the standard multi-temperature disc
black body \cite{mitsuda:84a} plus a power-law component.  We modeled
absorption in the intervening interstellar medium by fixing the equivalent
hydrogen column to the value found from radio observations $N_{\rm
  H}=3.2\times 10^{20}\,\rm cm^{-2}$ (Staveley-Smith 1999, priv.\ comm.),
and by using the cross sections of Ba\l{}uci\'{n}ska-Church \& McCammon
\shortcite{balu:92a}.  Our preliminary fits to this model had strong
residuals in the region around $\sim 6.5$\,keV (Fig.~\ref{fig:x3longspec},
middle).  We therefore included a narrow emission line at 6.4\,keV.  We will
return to the issue of the reality of this line feature below.
Table~\ref{tab:x3fits} lists the parameters of the best fit models.

The characteristic temperature of the phenomenological accretion disc
component, $kT_{\rm in}$, can be determined to high accuracy (see also
Paper~II). The formal uncertainty of the values for $kT_{\rm in}$ presented
in Table~\ref{tab:x3fits} are typically smaller than $0.005$\,keV. This
rather small formal uncertainty value might indicate that the systematic
uncertainty of the response matrix needs to be taken into account even for
these short observations.  These systematic uncertainties, however, are of
roughly the same magnitude as the Poisson errors used in our analysis. This
fact makes a formal error analysis rather difficult, and it is not clear
whether the parameter error analysis methods of Lampton et al.
\shortcite{lampton:76a} yield meaningful results in such a case.

During the week over which the observations were performed, $kT_{\rm in}$
remained more stable than the other fit parameters, although these other
parameters had relatively weak variations. It has been observed that on
much longer time scales (months to years) variations of \xthree\ are
correlated predominantly with strong variations of $kT_{\rm in}$, while the
phenomenological disc normalisation, $A_{\rm disc}$, remains relatively
constant (Paper~II, Ebisawa et al.  1993\nocite{ebisawa:93a}).  In the
literal interpretation of the multi-temperature disc black body, $A_{\rm
  disc}\propto r_{\rm in}^2 \cos i$, where $r_{\rm in}$ is the inner radius
of the accretion disc and $i$ is its inclination.  The stability of $A_{\rm
  disc}$ thus has been interpreted as evidence that the inner radius of the
accretion disc stays remarkably constant on long timescales
\cite{ebisawa:93a}. As discussed by Merloni et al. \shortcite{merloni:99a},
however, this is not necessarily true for a more physical disc model that
accounts for radiative transfer effects, Doppler blurring, and
gravitational redshifting.  For example, a model wherein only the accretion
rate and not the inner disc radius varies formally can be fit with a disc
black body with varying $A_{\rm disc}$ (Merloni et al. \nocite{merloni:99a}
1999).

\begin{figure}
\centerline{\includegraphics[width=0.45\textwidth]{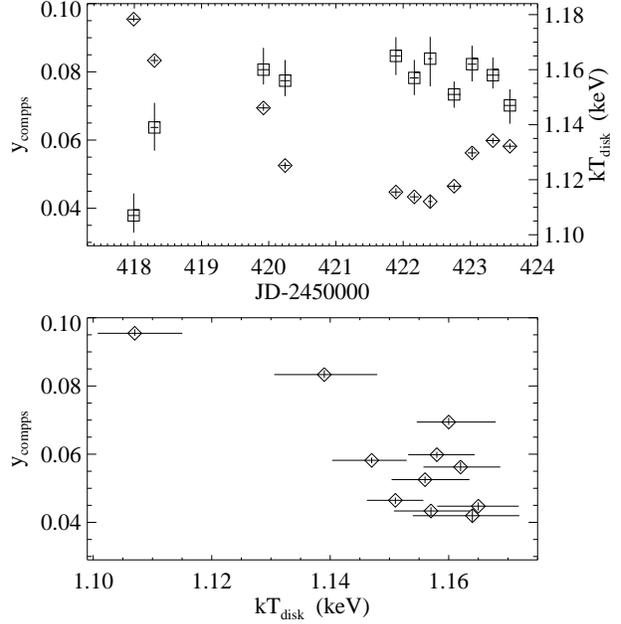}}
\caption{{\it Top:} Peak disc temperature and Compton $y$-parameter as a
  function time for the observations of \xthree\ presented in
  Tab.~\protect{\ref{tab:x3log}}. {\it Bottom:} Peak disc temperature vs.
  Compton $y$-parameter. \label{fig:dpcorr}}
\end{figure}

As a more physically motivated description, we have also fit the
Compton scattering model of Poutanen \& Svensson
\shortcite{poutanen:96a} ({\tt compps}). Specifically, we simulate a
disc black body Compton upscattered via a slab geometry corona with
fixed unity covering fraction over the disc.  A purely thermal coronal
electron distribution was assumed, and the width of the Fe line was
fixed to be $\sigma \sim 1.4$\,keV (see below). We then fit the peak
temperature of the disc black body, as well as the electron
temperature and Compton $y$-parameter of the corona. Results are
presented in Table~\ref{tab:x3log} and in Fig.~\ref{fig:dpcorr}.  Here
the peak temperature of the disc is seen to vary more strongly over
the course of the observation than for the multi-temperature black
body plus power-law models.  The variations in the lightcurve are
consistent with being predominantly driven by variations of the
Compton $y$-parameter, with the large count rate change from
observations a to e corresponding to a drop in $y$. The slight increase
in count rate seen in observations h--k corresponds to a slight
increase in $y$.  There is a tendency for the peak disc temperature and
Compton $y$-parameter to be anti-correlated; however, this
anti-correlation is dominated by observations a and b.

Note that the lightcurves cover approximately 3, 1.7\,day orbital periods
of \xthree.  Recent work with these (now archived) \rxte\ observations,
along with more recent long-term pointings, reveal evidence for a weak (few
percent) orbital modulation of the X-ray source if one folds the
lightcurves on the known orbital period (Boyd \& Smale, in prep.).
Such variations would not be unexpected as previous observations have
suggested a near edge-on system to account for the ellipticity of the
optical lightcurve \cite{vanderklis:83a}.  The nature of these variations,
possibly due to weak scattering, will be discussed further by Boyd \& Smale
(2000).

We now consider fits to the summed spectrum from the long observation.
Taking PCA channels 7 to 58 ($\approx 3.6$--26\,keV), we again fit two
models: a disc black body plus power law plus gaussian line, and a {\tt
  compps} model plus gaussian line.  The best fit disc black body plus power
law models are shown in Fig.~\ref{fig:x3longspec}.  For both classes of
models, we formally required a broad Fe line component with equivalent
width $\aproxgt 150$\,eV with widths of $\sigma \approx 1.4$\,keV
(Table~\ref{tab:x3fits}).  This is approximately twice the \pca\ spectral
resolution.  Despite the low reduced $\chi^2$, we do not consider such a
strong line to have been detected definitively in \xthree\ as there is some
concern about systematic errors in both the \pca\ response and the spectral
model.

Excluding the line, residuals in the line region are $\sim 1\%$,
similar to the systematic uncertainties in the \pca\ response. In
addition, for the models that we have fit, energy channels $\aproxlt
6$\,keV are dominated by the disc component, whereas energies
$\aproxgt 10$\,keV are dominated by the power-law/Comptonised
component.  The crossover region corresponds to the Fe line region
(see Fig.~\ref{fig:x3longspec}).  Thus the line strength is especially
sensitive to any errors in modelling the continuum spectral shape,
such as, e.g., the approximation made by the diskbb+power model, where
the comptonised spectrum is represented by a pure power-law even at
energies where the comptonised photons have energies comparably to the
seed photon energy. This is in contrast to AGN, for example, where the
continuum through the line/edge region is thought to be reasonably
well-approximated by a featureless power law, and therefore is not
subject to systematic uncertainties in the continuum model.

For the case of AGN, the reality of the line can also be verified
(independently of response matrix uncertainties) by taking the ratio
of the (nearly power-law) observed spectrum to that of the power-law
spectrum of the Crab nebula. In principle, a similar procedure could
also be used in case of \xthree.  Simulating a {\tt compps} spectrum
both with and without a line, utilizing the fit parameters for
\xthree\ discussed above, shows that a division of these two spectra
would reveal the presence of a line. There exists no template
spectrum, however, with which to do this comparison.  We caution,
therefore, that the \xthree\ line in reality may be weaker and/or
narrower than that obtained here.

The Comptonisation model implies a coronal temperature of $\approx
60$\,keV, and a Compton $y$-parameter of $y \equiv 4kT_{\rm es}/mc^2 ~
\tau_{\rm es} \approx 0.06$, which implies an optical depth of
$\tau_{\rm es} \approx 0.1$.  Due to the limited energy range of our
observations, however, these fits are not unique. We assumed a purely
thermal electron distribution, but we note that a purely non-thermal
Comptonizing electron distribution fits the data equally well. Data at
energies significantly higher than 50\,keV are required to break this
degeneracy.

As regards the power-law component, when considering the \hexte\ data as
well, we detect the source out to $\approx 50$\,keV, the highest energy at
which this source has ever been detected.  The \hexte\ data, however, do
not place much stronger constraints than were obtained with the \pca\ data
alone.  Including the \hexte\ data, the observed power law is $\Gamma =
2.8_{-0.2}^{+0.1}$.  We do not detect any noticeable curvature in the
spectrum, but our limits on the presence of such curvature are weak.

\subsection{\xone}
Historically, \xone\ has shown relatively little long time scale
variability (Syunyaev et al. 1990\nocite{sunyaev:90a}, see also Paper~II).
Consistent with this, \xone\ showed no obvious variations on day long
time scales during the long \rxte\ observation of 1996~December~6 to~8. The
combined spectrum is a typical example of the spectrum of \xone\ (see
Paper~II).  Taking our screening criteria into account, we obtained
80\,ksec of data with 5 PCU on, and another 30\,ksec of data wherein there
were 3 or 4 PCU on. Again, these data have been added together as for
\xthree.  This resulted in a detection of \xone\ out to 20\,keV in the
\pca. In \hexte, \xone\ is detectable out to the same energy, but due to
its low count rate (the total \hexte\ count rate is $\aproxlt 1\,\rm
counts\,s^{-1}$), no spectral information can be obtained.

\input{Table3}

\begin{figure}
\centerline{\includegraphics[width=0.5\textwidth]{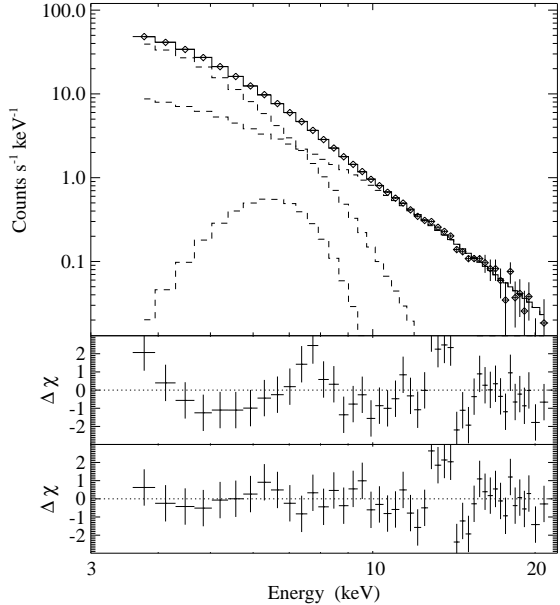}}
\caption{Same as Fig.~3 for the long observation of
  \xone. \label{fig:x1longspec}} 
\end{figure}

We modeled the data (PCA channels 7 to 52, $\approx 3.6$--22\,keV) with
the same spectral models as that for \xthree, using an equivalent hydrogen
column of $N_{\rm H}=7.2\times 10^{21}\,\rm cm^{-2}$ (Staveley-Smith 1999,
priv.\ comm.), taken from 21\,cm measurements.  Again, the data required
the presence of an iron line at 6.4\,keV. The resulting best fit
parameters, with a narrow and with a broad iron line, are shown in
Table~\ref{tab:x1fits}.  In Figure~\ref{fig:x1longspec}, we display the
residuals of the disc black body fits, showing the improvement of assuming a
broad iron line. As for \xthree, without a broad line the relative
deviation between the data and the model in the line region is $\approx
1\%$.  Again, the line region corresponds to the transition region between
the disc black body and power-law/Comptonized emission components of the
spectrum, so caution is warranted in interpretting the fit parameters of
the line.

Our spectral parameter values are in general agreement with earlier data
\cite{ebisawa:89a,schlegel:94a}, although our power-law index appears to be
rather soft compared to the earlier measurements.  As we show in
Fig.~\ref{fig:x1longspec}, the soft ($\Gamma\sim 3$) power-law component
dominates the observed flux above $\sim$7\,keV, but also contributes an
appreciable fraction of the total soft flux. Thus the lower energy channels
are also in part determining the value of the fitted spectral index.  In
terms of the {\tt compps} Comptonisation model, if we choose a purely
thermal, slab-geometry corona with unity covering factor, we find similar
coronal parameters as for \xthree; however, the disc temperature is
somewhat lower so the observed spectrum is more dominated by Comptonized
photons than that for \xthree.  Again, purely non-thermal models are also
permitted, indicating that the coronal geometry and parameters are not
uniquely determined by these data.

Since our observations cover a larger energy range than many previous
missions, they ensure that our best fit models at least describe the
spectral shape quite well.  In accordance with Schlegel et al.
\shortcite{schlegel:94a}, we do not find evidence for a broad iron edge
feature in the data.  Such a feature might be artificially introduced in
the data for instruments where the upper energy boundary is so low that the
power-law index cannot be well constrained.

\section{Archival ASCA Data}\label{sec:asca}

The High Energy Astrophysics Archive (HEASARC) contains one \asca\ 
observation of \xone\ (1995 April 2) and two observations of \xthree\ (1993
September 23, 1995 April 15). To the best of our knowledge, these
observations have not been published previously.  We extracted the data
from all four instruments on \asca, the two solid state detectors (\sis0 and
\sis1) and the two \gis\ detectors (\gis2 and \gis3); however, here we only
discuss the \sis\ data, as the \sis\ has a more reliable low energy response 
and a better spectral resolution.

We used the SISCLEAN tools \cite{day:98a}, with the default values, to
remove hot and flickering pixels.  Furthermore, we filtered the data with
the same cleaning criteria outlined by Brandt et al.
\shortcite{brandt:96a}; however, we took the more conservative values of
$10^\circ$ for the minimum elevation angle and 7\,$\mbox{GeV}/c$ for the
rigidity.  We obtained background estimates by extracting source-free
regions near the edges of the \sis\ chips.  The spectra were rebinned so that
each energy bin contained a minimum of 20\,photons, and we only fit \sis\
data in the 0.5 to 10\,keV range.  Cross-calibration uncertainties between
the two \sis\ detectors were accounted for by introducing multiplicative
constants (always found to be within $\aproxlt 1\%$ of each other) for each
detector in all fits.

\input{Table4}

\begin{figure}
\centerline{
\resizebox{\hsize}{!}{\includegraphics{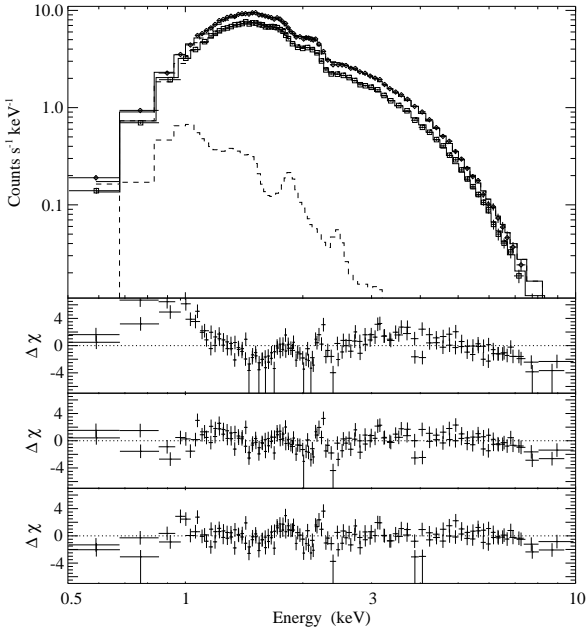}}
}
\caption{\small \asca\ observations of \xone. Top panel shows the {\tt
    compps} plus a Raymond-Smith plasma model fit to the data, with $N_{\rm
    H}$ fixed to $7.2\times10^21~{\rm cm^{-2}}$.  Dashed lines show the
  individual model components folded through the response matrix of the
  \sis0 detector.  Lower panels, from top to bottom, show data residuals
  for: {\tt compps} model with fixed $N_{\rm H}$ and no plasma component,
  {\tt compps} model with fixed $N_{\rm H}$ and a Raymond-Smith plasma
  component, and {\tt compps} model with free $N_{\rm H}$ and no plasma
  component.  \protect{\label{fig:lmcx1}}}
\end{figure}

Neutral hydrogen absorption of a disc black body spectrum \cite{mitsuda:84a}
provides a rough description of the \asca\ data.  Such models yielded
reduced $\chi^2$ values $\aproxgt 2$ for \xone\ and $\approx 1$ for
\xthree.  All three observations, however, exhibited evidence for excess
emission at energies $\aproxgt 5$\,keV.  Adding a power law (photon flux
$\propto E^{-\Gamma}$, with $\Gamma \approx 2.2$--2.5) improved the reduced
$\chi^2$ to $\approx 1.1$ and $\approx 0.7$ for \xone\ and \xthree,
respectively.  As for the \rxte\ observations, we also were able to fit the
{\tt compps} model; however, given that the \asca\ data cuts-off at 10\,keV
there was not enough leverage to fit both $kT_{\rm e}$ and $y_{\rm
  compps}$. We therefore froze the coronal electron temperature to 50\,keV
for all three observations.  Results for these fits are presented in
Table~\ref{tab:ascafit} and Fig.~\ref{fig:lmcx1}.

Both \xthree\ observations are at very low flux values. We compared the
3-9\,keV flux values from these observations to those obtained from our
\rxte\ monitoring observations in Paper~II.  Using an \sis/\pca\ 
normalisation ratio of 0.7, we estimate that the \asca\ observations would
correspond to \asm\ count rates of $\sim 1.3$\,cps and $\sim 0.6$\,cps.  As
discussed in Paper~II, these low count rates occur near transitions from
the soft to hard state of \xthree. We note that the lower flux \asca\ 
observation of \xthree\ shows a significant drop in peak disc temperature,
consistent with the trends seen in the state transitions discussed in
Paper~II.

Our main purpose for examining the \asca\ data is to determine if there is
evidence for an Fe line in the spectrum and an Fe-L complex near energies
of $\approx 1$\,keV. Line features near 1\,keV are a common occurrence in
photoionized plasmas close to sources emitting hard X-rays (e.g., in
eclipse in Vela~X-1, Nagase et al. 1994\nocite{nagase:94a}). We would also
expect such features in models of warped accretion discs with winds,
similar to those of Schandl \shortcite{schandl:96a}, or in models of
wind-driven limit cycles (Shields et al. \nocite{shields:86a} 1986), as
might be relevant for producing the long term periodicity of LMC~X-3 (Paper
II).  Such a line complex also was seen in the spectrally very similar
source V1408~Aql$=$4U1957+11 \cite{nowak:99d}.

Starting with the {\tt compps} models, none of the three \asca\ 
observations show evidence of an Fe line. If we freeze the line energy at
6.4\,keV or 6.6\,keV and the line width at $0.3$\,keV, we find an upper
(90\% confidence) limit of 60\,eV for the line equivalent width for the
\xone\ observation. Narrower lines (down to the \asca\ resolution of $\sim
0.1$\,keV) produce more stringent equivalent width limits.  For the
slightly fainter 1995 April 15 and 1993 September 23 \xthree\ observations,
the equivalent width 90\% confidence limits are 90\,eV and 320\,eV,
respectively.

The reduced $\chi^2$ values with the {\tt compps} model are significantly
less than one for both \xthree\ observations, and therefore we do not
attempt to fit a plasma component to these data.  The {\tt compps} model
fit to the \xone\ observation yields a reduced $\chi^2 \approx 1$; however,
this fit requires an $N_{\rm H}$ value 12\% lower than the estimates of
Staveley-Smith (1999, priv. comm.). Fixing the $N_{\rm H}$ column density
to this value worsens the reduced $\chi^2$ to $\approx 2$ and yields
significant residuals in the Fe-L complex region (Fig.~\ref{fig:lmcx1}).
This residual can be removed and the reduced $\chi^2$ value returned to
$\approx 1$ by adding a plasma component with $kT \sim 0.8$\,keV (here
modeled with the XSPEC {\tt raymond} model, after Raymond \& Smith
\nocite{raymond:77a} 1977). Thus the reality of any Fe-L complex is seen to
depend strongly upon accurate modeling of the neutral hydrogen absorption
of this source.  Differing metal abundances in the LMC also could lead to
errors in interstellar absorption models, adding further uncertainties to
the level of any plasma component.  If real, however, a plasma component at
the levels allowed by the \asca\ data with a fixed neutral hydrogen column
would be readily detectable by the X-ray Multiple Mirror Mission, which
would also resolve the complex into individual line components (Note
also that the residuals seen at $\sim 1.7$ and $\sim 2.2$\,keV are
likely related to well-known systematic features in the \sis\ responses
matrices.)

\section{Timing Analysis}\label{sec:timing}

\begin{figure}
\centerline{
\resizebox{0.5\textwidth}{!}{\includegraphics{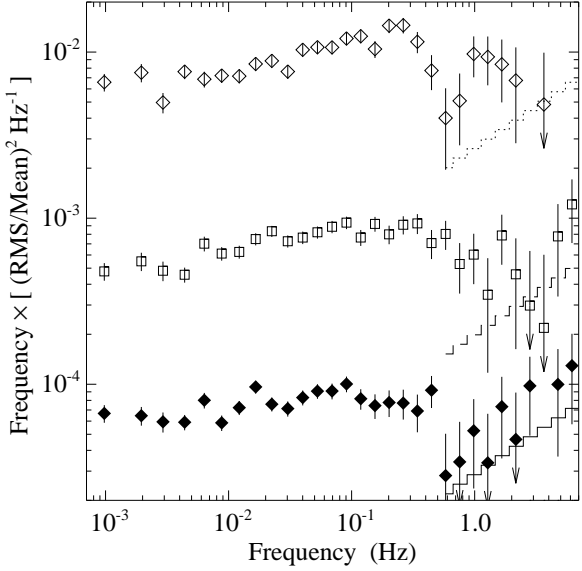}}
}
\caption{\small Fourier frequency times Power Spectral Density
  (PSD) for \xone\ in the 0--3.3\,keV (solid diamonds, lowered by a
  decade), 3.3--4.7\,keV (clear diamonds), 4.7--9.1\,keV (solid squares,
  raised by a decade) energy bands.  PSD normalisation is such that
  integrating over positive frequencies yields the mean square variability
  divided by the square of the mean for the lightcurve analyzed. Lines
  represent the expected level of positive 1-$\sigma$ noise residuals after
  subtracting Poisson noise from the PSDs.  \protect{\label{fig:psd}}}
\end{figure}

We employed Fourier techniques, in the same manner as for our RXTE
observations of Cyg X-1 \cite{nowak:99a}, to study the short timescale
variability of \xone\ and \xthree. Specifically, we used the same
techniques for estimating deadtime corrections \cite{zhangw:95a,zhangw:96a}
to the Power Spectral Density (PSD), and for estimating uncertainties and
the Poisson noise levels of the PSD \cite{leahy:83a,vanderklis:89b}.  In
the analysis discussed below, we use lightcurves with $2^{-7}$\,s
resolution constructed from the \pca\ top Xenon layer data only.
Furthermore, we subdivide the lightcurves into three energy channels:
0--3.3\,keV, 3.3--4.7\,keV, and 4.7--9.1\,keV (absolute \pca\ channels
0--8, 9--12, 13--24, respectively).  We chose these energy ranges as they
are relatively background free, and furthermore the three channels have
roughly equal count rates.  The lowest energy channel is dominated by the
disc component, whereas the highest energy channel is sampling the power
law component (\S\ref{sec:spectra}).

To maximize the signal to noise ratio, we created a single PSD in each
energy channel that was averaged over the entire duration of our
observations.  We further averaged PSDs constructed from data segments of
1024\,sec duration (76 segments for \xone, 65 segments for \xthree), and we
logarithmically binned the PSD over frequencies $f \rightarrow 1.3f$.  For
the case of \xthree, weak variability with root mean square (rms) amplitude
of 0.8\% of the mean was detected in the $10^{-3}$--$10^{-2}$\,Hz range.
This is consistent with background fluctuations.

Significant variability above the noise, however, was detected in \xone.
Root mean square variabilities of 6.5\%, 6.3\%, and 7.1\%, lowest to
highest energy channel, were found for the frequency range
$7\times10^{-4}$--0.3\,Hz. The PSDs for these energy channels are shown in
Fig.~\ref{fig:psd}. All three PSDs are approximately proportional to
$f^{-1}$ from $7\times10^{-4}$--0.2\,Hz.  The highest energy band PSD,
however, shows evidence for a rollover at $\sim 0.2$\,Hz.  Note that for
this latter PSD there are only three or four points at frequencies higher
than 0.2\,Hz that sharply decline, followed by several points with a more
gradual decline roughly proportional to $f^{-0.5}$.  (Fig.~\ref{fig:psd}
shows the PSD multiplied by frequency.)  The $f^{-0.5}$ behaviour is also
readily apparent at high frequencies in the PSD of the low energy channel
as well as possibly in the PSD of the middle energy channel.  An $f^{-0.5}$
proportionality is exactly that expected for positive noise residuals if
the mean Poisson noise level was slightly underestimated (see Nowak et al.
1999a\nocite{nowak:99a}).  The uncertainty in the noise level makes it
difficult to assess the frequency-dependence of the rollover as a function
of energy. The 4.7--9.1\,keV energy band PSD values at frequencies
0.2--2\,Hz, however, clearly lie below an extrapolation of the
low-frequency PSD behaviour.  Removing the break altogether would require
that we \emph{overestimated} the Poisson noise level, which seems doubtful
given the $\propto f^{-0.5}$ behaviour of the noise-subtracted PSD
residuals.

These results for \xone\ are roughly consistent with the previous \rxte\ 
observations reported by Schmidtke et al. \shortcite{schmidtke:99a} in
terms of overall PSD amplitude and shape.  However, Schmidtke et al.
\shortcite{schmidtke:99a} did not attempt any noise subtraction and did not
search for breaks in any of the PSD.  We do not find any evidence for a
0.08\,Hz quasi-periodic oscillation as reported by Ebisawa et al.
\shortcite{ebisawa:89a}, which is consistent with the results of Schmidtke
et al. \shortcite{schmidtke:99a}.  The 0.08\,Hz QPO reported by Ebisawa et
al.  \shortcite{ebisawa:89a}, however, is nearly coincident with the
expected level of the residuals after noise subtraction.  Thus the
previously reported QPO may have be an artifact of a misestimation of the
Poisson noise level, as opposed to the lack of detection here and in the
work of Schmidtke et al.  \shortcite{schmidtke:99a} being due to an
intermitant nature of such low-frequency QPO.

Contrary to the results reported by Treves et al. \shortcite{treves:88a},
we do not detect any variability from \xthree.  We note, however, that the
PSD reported by Treves et al.  \shortcite{treves:88a} has the
characteristic $f^{-0.5}$ shape and amplitude expected from noise
residuals. We believe, therefore, that the results of Treves et al.
\shortcite{treves:88a} are consistent with those reported here.

\begin{figure*}
\centerline{
\resizebox{0.33\hsize}{!}{\includegraphics{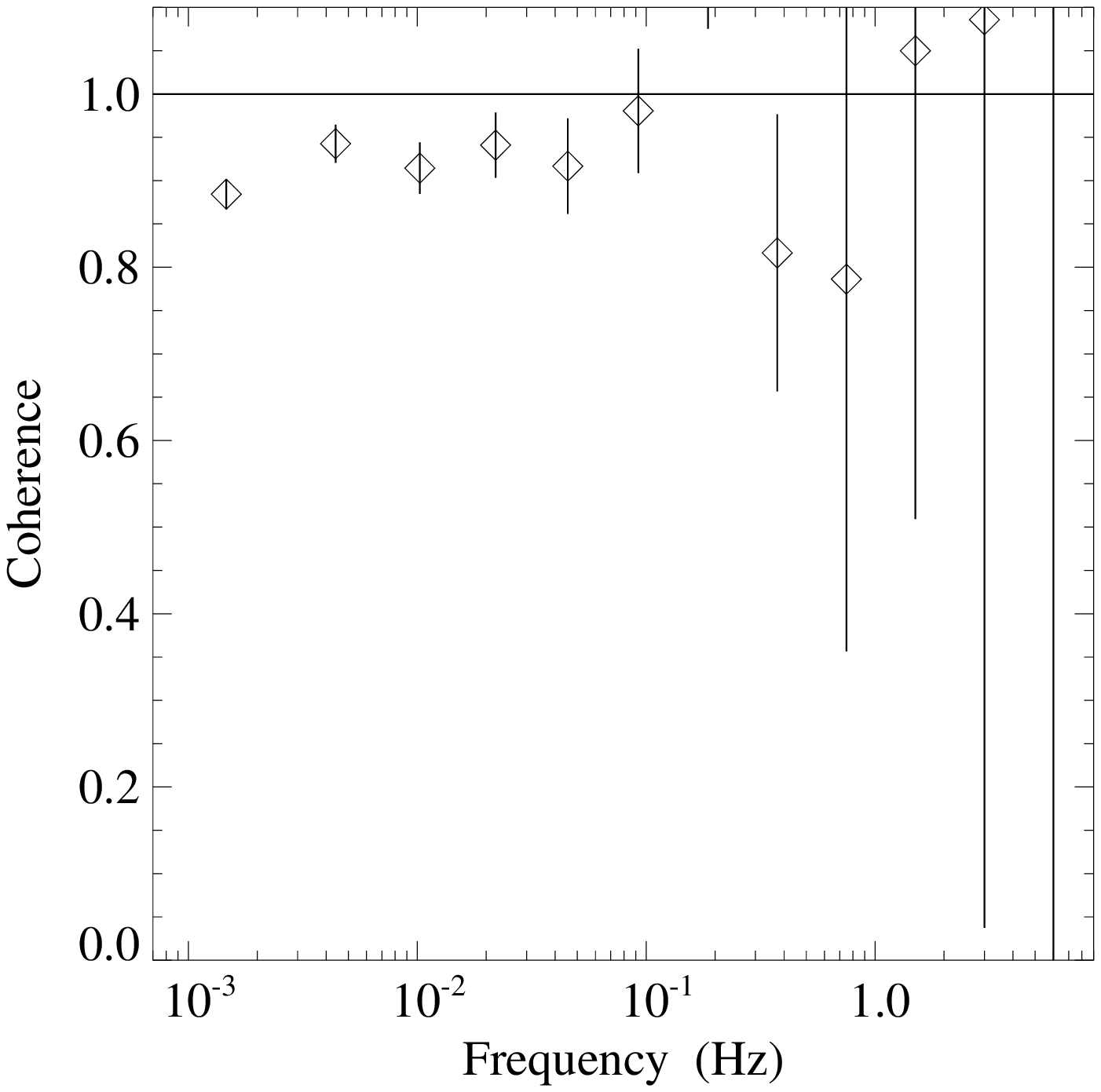}}
\resizebox{0.33\hsize}{!}{\includegraphics{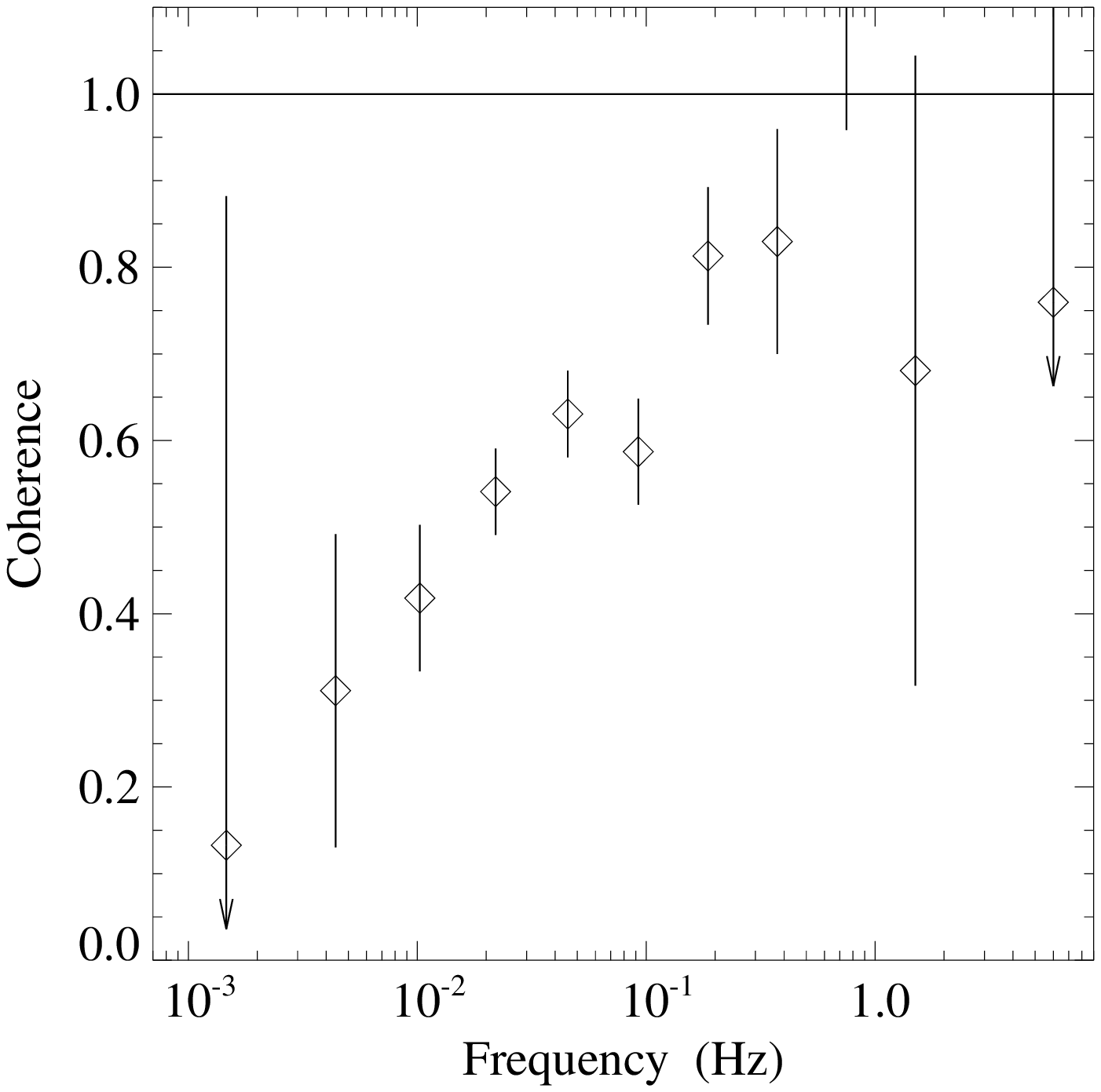}}
\resizebox{0.33\hsize}{!}{\includegraphics{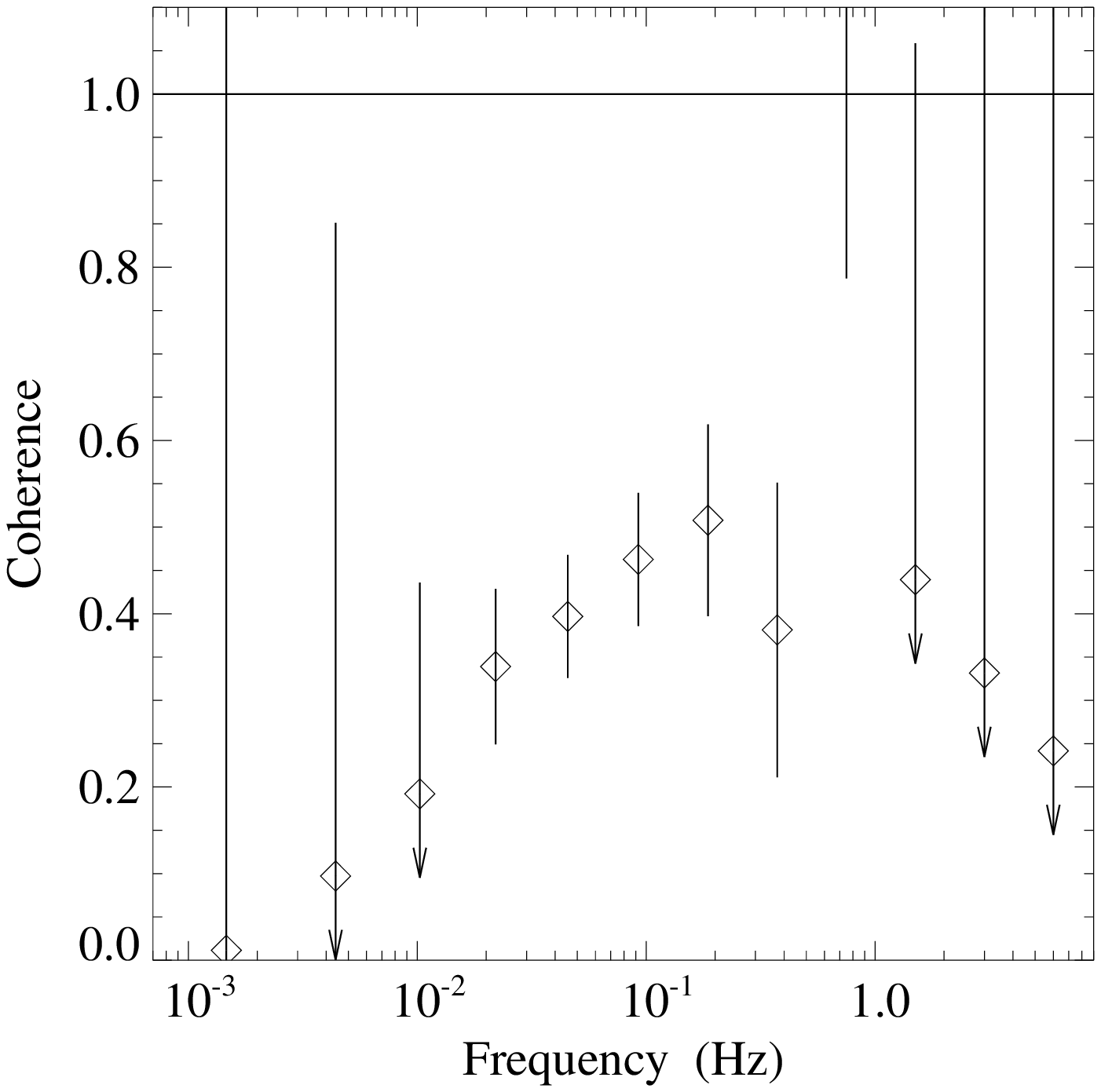}}
}
\caption{\small
  Coherence between the variability in the 0--3.3\,keV and 3.3--4.7\,keV
  energy bands (left), the 3.3--4.7\,keV and 4.7--9.1\,keV energy bands
  (middle), and the 0--3.3\,keV and 4.7--9.1\,keV energy bands (right).
  \protect{\label{fig:coh}}}
\end{figure*}

We further studied the temporal behaviour of \xone\ by computing the
Fourier frequency-dependent time lags and variability coherence between the
various energy bands (see Vaughan \& Nowak 1997\nocite{vaughan:97a}, and
references therein).  Due to the low count rates, only weak 1-$\sigma$ upper
limits of $\aproxgt 2~{\rm sec}~ (f/10^{-2}~{\rm Hz})^{-1}$ could be placed
on the time lags between the lowest and highest energy channels over the
frequency range $10^{-3}$--$0.3$\,Hz.  We were able to measure the
coherence among the variability in the various energy bands. As we have
previously discussed (Vaughan \& Nowak 1997\nocite{vaughan:97a}; see also
Bendat \& Piersol 1986\nocite{bendat}), the coherence function,
$\gamma^2(f)$, which is always $\le 1$, is a measure of the degree of linear
correlation between two time series, or alternatively, it is a measure of
how well one time series can be predicted from the other.

We present the coherence function for the \xone\ variability lightcurve,
logarithmically binned over frequencies $f \rightarrow 2f$, in
Fig.~\ref{fig:coh}.  The lowest and middle energy band lightcurves are
fairly well-correlated and have a coherence function between them of
$\gamma^2(f) \aproxgt 0.9$. Both of these lightcurves, however, are very
incoherent with the highest energy band lightcurve, with $\gamma^2(f)
\approx 0$ at $10^{-3}$\,Hz and then rising approximately as $\propto
\log(f)$. Due to uncertainties in the Poisson noise level there are
additional systematic uncertainties above $\approx 0.3$\,Hz; however, the
coherence is clearly below the near unity levels that we typically observe
during the low/hard states of BHC such as Cyg~X-1 and GX339-4
\cite{nowak:99a,nowak:99c}.

\section{Discussion}\label{sec:discuss}

\subsection{Variability of \xone} \label{sec:xonevari}

The most obvious striking difference between \xone\ and \xthree\ is that
the former shows moderate (7\% rms) variability whereas the latter shows
essentially none ($< 0.8\%$ rms) on timescales shorter than 1\,ks, despite
the fact that the X-ray spectra of these two objects are remarkably similar
in terms of shape and overall flux.  We hypothesize that the differences
lie within their respective modes of accretion.  As discussed by Cowley et
al. \shortcite{cowley:83a}, van der Klis et al. \shortcite{vanderklis:83a},
and Cowley et al. \shortcite{cowley:94a}, \xthree\ is consistent with
accreting via Roche lobe overflow.  One would therefore expect the
accretion rate of \xthree\ to be relatively steady (although see the
discussion of the long-term periodicity presented in Paper II). \xone, on
the other hand, is most likely accreting partly via a wind with velocity
$v_{\rm w} \aproxgt 600$--$1100~{\rm km~s^{-1}}$
\cite{hutchings:83a,hutchings:87a}.  This has a number of implications for
the accretion flow.

Taking a presumed mass of $6~\Msun$ and a wind velocity of $600~{\rm
  km~s^{-1}}$ \cite{hutchings:87a}, the Hoyle-Lyttleton accretion radius
  \cite{hoyle:39a,bondi:44a} is then
\begin{equation}
R_{\rm HL} = \frac{2 GM}{v_{\rm w}^2} \approx 4 \times 10^{11}~{\rm cm}
~\left(\frac{M}{6\,\Msun}\right) ~\left(\frac{v_{\rm w}}{600 ~{\rm
      km\,s^{-1}}}\right)^{-2} ~~,
\label{eq:hoyle}
\end{equation}
which should be compared to the binary separation of
$a\sim2.2\times10^{12}~{\rm cm}$ given a 4.2\,day period and a $20~\Msun$
companion \cite{hutchings:87a}. The ``circularisation radius'', however,
will be much smaller, by a factor of the order $(v_{\rm orb}/v_{\rm w})^2$,
where $v_{\rm orb} \sim 260~{\rm km~s^{-1}}$ is the orbital velocity. This
estimate comes from assuming the accreted angular momentum per unit mass
goes as $R_{\rm HL} \cdot v_{\rm orb}$, and then setting this value equal
to the Keplerian value at the circularisation radius.  Thus, we expect a
disc circularisation radius of $\aproxlt 8\times10^{10}~{\rm cm}~(v_{\rm
  w}/600~{\rm km~s^{-1}})^{-4}$.  The disc is thus expected to be smaller
than would be usual for Roche lobe overflow, perhaps substantially so.  For
$v_{\rm w} = 2000~{\rm km~s^{-1}}$, the expected circularisation radius is
$\sim 750~GM/c^2$.  

There is a long history of numerical simulations
\cite{fryxell:88a,taam:88a,taam:89a,benensohn:97a,ruffert:97a,ruffert:99a}
showing that wind dominated accretion inherently exhibits fluctuations in
the net accretion rate onto the compact object.  More recent 3-D numerical
integrations \cite{ruffert:97a,ruffert:99a} tend to show smaller amplitude
fluctuations than earlier 2-D work \cite{benensohn:97a}; however, accretion
rate fluctuations of ${\cal O}(10\%)$, in rough agreement with the observed
variability, are expected. \xthree, being Roche lobe fed, would not be
expected to exhibit these fluctuations.

A disc with a truncated outer edge may also yield an explanation of the
possible rollover in the high-energy band PSD seen at $\sim 0.3$\,Hz.  For
$\alpha=0.3$, $H/R \sim L/L_{\rm Edd} \sim 0.3$, $M=6~\Msun$, the viscous
timescale at $R=750~{GM/c^2}$ is $\sim 20$\,sec, which is in rough
agreement with the time scale of the possible rollover.  This is consistent
with the idea that any accretion rate fluctuations on time scales shorter
than the viscous time scale at the circularisation radius are smoothed out
as they propagate into the inner X-ray emitting regions of the disc.

As discussed above, the low variability coherence values for \xone\ at $f >
10^{-3}$\,Hz are unlike what we observed for the low/hard states of Cyg~X-1
and GX~339$-$4 \cite{nowak:99a,nowak:99c}.  We are aware of three possible
ways of explaining such low values \cite{vaughan:97a}.  First, the low and
high energy bands may be completely unrelated to each other. This seems
unlikely as all three energy bands exhibit a roughly $\propto f^{-1}$ PSD
with comparable amplitudes.  Second, the upper energy band PSD may be
related to the lower energy band PSD via a non-linear transfer function.
This is a viable option, given the complicated physics of ``coronal
formation'' and the fact that non-linear transfer functions couple power
from frequencies $f$ in one band to multiples of $f$ in other bands
\cite{bendat}. This latter case is especially difficult to detect in the
$f^{-1}$ PSD because of its scale-free nature.  

The third possibility is perhaps the most viable option. ``Mixing'' of
\emph{independent} components with widely varying intrinsic time lags can
lead to the loss of coherence (see the discussion presented by Nowak et al.
1999b\nocite{nowak:99c}).  For example, if one component which shows a
correlation between hard and soft X-rays (e.g., local dynamical time scale
disc fluctuations leading to and increase in both seed photons and, very
slightly delayed, upscattered hard photons) is mixed with another, nearly
equal in power \emph{but independent} component with an anti-correlation
between soft and hard X-rays (e.g., matter being transferred between disc
and corona on thermal or viscous time scales, with comparably long delays
between soft and hard photons), then a strong loss of coherence is
expected.  In a simple two-component model the coherence function,
$\gamma^2(f)$, is given by
\begin{equation}
\gamma^2(f) \approx \frac{ R^2(f) + 2 R(f) \cos ( \theta_b - \theta_a ) }
     {[ R(f) + 1 ]^2} ~~.
\end{equation}
Here $R(f)$ is the ratio of Fourier frequency-dependent cross correlation
amplitudes for the two independent processes being considered (soft vs.
hard X-ray variability, process $b$ over process $a$), and $\theta_a$,
$\theta_b$ are the Fourier frequency-dependent phase delays between the
hard and soft X-ray variability for the independent processes $a$ and $b$,
respectively \cite{vaughan:97a,nowak:99c}.  (Time delay, $\tau \equiv
\theta/2\pi f$.) For the plots in Fig.~\ref{fig:coh}, if $R(f)\approx 1$
and $\theta_b - \theta_a \approx \pi$ at $f = 10^{-3}$\,Hz, then
$\gamma^2(f) \approx 0$ there as well.  The increase in coherence towards
higher frequency could be due to a decrease of either $\theta_b - \theta_a$
or $R(f)$.  The latter might be reasonably expected, for instance, if one
process is happening predominantly on ${\cal O}(1000~{\rm sec})$ viscous
time scales (again, perhaps matter/energy transfer between disc and
corona), whereas the other one is happening over a broader range of time
scales (fluctuations in the seed/Comptonized photons).

\subsection{Spectroscopy of the Soft State} \label{sec:soft}

The spectroscopy of \xthree\ perhaps bears out the notion of there being an
anti-correlation between the disc and coronal components on long time
scales, as shown in Fig.~\ref{fig:dpcorr}.  A similar result can be
obtained with the more phenomenological disc black body plus power law
models presented in Table~2.  The photon flux in the `disc component'
($\propto A_{\rm disc} kT_{\rm disc}^3$) is anti-correlated with the photon
flux in the `power law component' ($\propto 3.6^{1-\gamma}/[1-\gamma]$)
in an analogous way.  The fastest variation occurs on time scales of
$\approx 0.4$\,days.  Given $\alpha = 0.3$, $H/R \sim L/L{\rm Edd} \sim
0.3$, and $M=9~{\rm M}_\odot$, this corresponds to a viscous time scale at a
radius of $\sim 8 \times 10^{10}$\,cm, approximately one quarter of the
disc circularisation radius.  It is therefore more likely that this
variation occurs on a thermal or dynamical time scale closer towards the
center of the system. Thermal time scale transfer of energy between disc
and corona is again consistent with the results shown in
Fig.~\ref{fig:dpcorr}. 

Unfortunately, it is difficult to say much more about the spectrum other
than it is well-fit by a disc black body plus power law model. Given a
$70^\circ$ inclination, $9~{\rm M}_\odot$ black hole, and 50\,kpc distance
for \xthree, and a $30^\circ$ inclination, $6~{\rm M}_\odot$ black hole,
and 50\,kpc distance for \xone, the disc normalisations in
Tables~\ref{tab:x3fits} and \ref{tab:x1fits} are roughly consistent with
inner disc radii of $\sim 6~GM/c^2$.  This is quite remarkable, given the
simplicity of the disc black body model (cf., Merloni et al.
\nocite{merloni:99a} 1999). As regards the power law tail, other soft state
black hole candidates exhibit evidence for non-thermal Comptonisation in
this tail \cite{gierlinski:99a}.  Here, however, as we were only able to go
out to 50\,keV and 20\,keV for \xthree\ and \xone, respectively, we were
able to describe the data completely adequately with pure thermal
Comptonisation.  

The extent to which a strong, broad Fe line is required is likewise still
somewhat uncertain.  The $\approx 150$\,eV equivalent width, $\sigma
\approx 1.4$\,keV line is consistent with expectations from the basic
picture outlined above: a disc extending down towards the black hole
horizon, with a large covering fraction corona sitting on top of it.  (A
slight Fe overabundance, or perhaps beaming of the coronal radiation
towards the disc might also be required for the implied equivalent widths.)
However, as the Fe line occurs in a region where the two basic spectral
components --- soft disc spectrum and hard power law/Comptonisation
spectrum --- cross one another, the Fe line parameters are especially
sensitive to our assumptions regarding the underlying continuum model.
Again, this is in contrast to AGN where it is presumed that the underlying
continuum in the Fe line/edge region is well-approximated by a simple power
law.

\section{Summary}\label{sec:summary}

In this work, we have presented analysis of long (170\,ksec) \rxte\ 
observations of \xone\ and \xthree, as well as analysis of archival \asca\ 
observations of these objects. The primary results of these analyses are as
follows.
\begin{itemize}
\item Both \xone\ and \xthree\ are well-fit by a simple spectroscopic model
  consisting of a disc black body and a power law. The power law, which was
  seen out to 50\,keV for \xthree, shows no obvious curvature.  The data
  are consistent with exhibiting a broad Fe K$\alpha$ line with equivalent
  width $\aproxlt 150$\,eV; however, the line could in reality be narrower
  and weaker. \asca\ observations do not reveal the presence of a narrow Fe
  K$\alpha$ line, and the presence of any Fe L-complex in \xone\ is almost
  completely dependent upon the modeling of the neutral hydrogen column
  towards that source.
\item As a more physical interpretation of these results, the spectra are
  consistent with Comptonisation of a disc spectrum by a corona with
  electron temperature $kT_{\rm e} \sim 60$\,keV and $\tau_{\rm es} \sim
  0.1$.  These fits are not unique, however, as the data do not extend
  beyond 50\,keV. Purely non-thermal Comptonisation models are also
  permitted. 
\item \xthree\ shows no variability on $\aproxlt 1$\,ks time scales;
  however, it does show variability on half day time scales and longer. The
  time scale of these variations are slightly faster than the viscous time
  scales of the outer disc.  Spectral models of these variations are
  consistent with an anti-correlation between the disc and Compton corona
  components of the spectrum
\item \xone\ shows rms variability of a few percent on faster than 1000\,s
  timescales.  This variability, along with with a possible power spectrum
  rollover on time scales $\aproxlt 5$\,s, is consistent with accretion
  rate variations in a wind fed system where the disc is circularized at
  radii $\aproxlt 1000~GM/c^2$.
\item The variability at energies $\aproxlt 5$\,keV shows very low
  coherence with the variability at energies $\aproxgt 5$\,keV.  Given the
  the similarity of the PSD among the three energy bands, this low
  coherence is likely either due to a non-linear relationship between the
  soft and hard energy bands, or due to multiple variability components
  some with strong correlations between soft and hard variability and other
  others with strong anti-correlations.
\end{itemize}

\section*{Acknowledgements}
This research has been financed by NASA grants NAG5-3225, NAG5-4621,
NAG5-4737, NSF grants AST95-29170, AST98-76887, DFG grant Sta 172/22,
and a travel grant to J.W.  and K.P. from the DAAD.  L. Staveley-Smith
provided us with the unpublished $N_{\rm H}$ values.  MAN and JW would
like to thank the hospitality of the Aspen Center for Physics and the
participants of the ``X-ray Probes of Relativistic Astrophysics''
workshop, especially J. Grindlay and R Taam, for many useful
discussions while this work was being completed. We thank the referee
for comments that improved the clarity of this paper.

\end{document}

%% file: Table1.tex
\begin{table*}
\caption{Observing log for the long RXTE observation of LMC~X-3. 
\protect{\label{tab:x3log}}}
\begin{center}
\begin{tabular}{ccccc}
%\hline
%
\noalign{\vskip 2pt}
{Obs.} & \multicolumn{2}{c}{Date} & {Exposure} 
& {Count Rate} \\ 
 & {JD-2450000} & {ymd}  & {sec}  & {counts\,s$^{-1}$} \\
\noalign{\vskip 2pt}
%
%\hline
%
\noalign{\vskip 2pt}
  a &  417.905 & 1996.11.30, 09:43 & 13800 &  $ 399.3\pm0.2$ \\ 
\noalign{\vskip 2pt}
  b &  418.238 & 1996.11.30, 17:42 & 10200 &  $ 412.9\pm0.2$ \\ 
\noalign{\vskip 2pt}
  c &  419.843 & 1996.12.02, 08:14 & 14000 &  $ 422.3\pm0.2$ \\ 
\noalign{\vskip 2pt}
  d &  420.168 & 1996.12.02, 16:02 & 12700 &  $ 411.3\pm0.2$ \\ 
\noalign{\vskip 2pt}
  e &  421.818 & 1996.12.04, 07:38 & 12400 &  $ 369.9\pm0.2$ \\ 
\noalign{\vskip 2pt}
  f &  422.100 & 1996.12.04, 14:25 & 11300 &  $ 388.8\pm0.2$ \\ 
\noalign{\vskip 2pt}
  g &  422.374 & 1996.12.04, 20:58 & ~5500 &  $ 384.2\pm0.3$ \\ 
\noalign{\vskip 2pt}
  h &  422.660 & 1996.12.05, 03:50 & 16900 &  $ 374.2\pm0.2$ \\ 
\noalign{\vskip 2pt}
  i &  422.960 & 1996.12.05, 11:02 & 11300 &  $ 334.9\pm0.2$ \\ 
\noalign{\vskip 2pt}
  j &  423.238 & 1996.12.05, 17:42 & 16400 &  $ 395.0\pm0.2$ \\ 
\noalign{\vskip 2pt}
  k &  423.502 & 1996.12.06, 00:02 & 15200 &  $ 392.2\pm0.2$ \\ 
\noalign{\vskip 2pt}
%
%\hline
\end{tabular}
\end{center}

\medskip
{Exposure times shown are rounded to the closest
100\,sec. The count rate is the total PCA background subtracted count
rate.}
\end{table*}

%%% Local Variables: 
%%% mode: latex
%%% TeX-master: t
%%% End: 

%% file: Table2.tex
\begin{table*}
\caption{Results of Spectral Fitting to the LMC X-3
  Data. \protect{\label{tab:x3fits}}} 
\begin{center}
\begin{tabular}{cllllllllcr}
%\hline
%
\noalign{\vskip 2pt}
{Obs.} & {$kT_{\rm in}$} & ${A_{\rm disc}}$ & $A_{\rm compps}$ & {$\Gamma$}
 & {$A_{\rm PL}$} & $kT_{\rm e}$ & $y_{\rm compps}$ & {$A_{\rm Line}$} &
 {EW} &  {$\chi^2/\rm dof$} \\ 
 & {keV} & & $10^{-2}$ & & {$10^{-1}$} & keV & $10^{-2}$ & {$10^{-4}$} &
 {eV} & \\ 
\noalign{\vskip 2pt}
%
%\hline
%
\noalign{\vskip 2pt}
total & $  1.24^{+0.00}_{-0.01}$ & $  33.6^{+0.9}_{-0.9}$ & $\ldots$ & $
2.0^{+0.1}_{-0.2}$ & $ 2.4^{+ 1.1}_{-  0.8}$ & $\ldots$ & $\ldots$ & $
0.6^{+ 0.5}_{- 0.6}$ & $  10 $ & $ 50.4/47$ \\ 
\noalign{\vskip 2pt}
total $^{\rm a}$ 
& $  1.20^{+ 0.01}_{- 0.02}$ & $  37.3^{+  1.8}_{-  1.4}$ & $\ldots$ & $
2.6^{+ 0.2}_{-  0.3}$ & $   1.2^{+ 0.7}_{-  0.6}$ & $\ldots$ & $\ldots$ & $
8.1^{+ 3.5}_{- 3.8}$ & $ 134 $ & $ 37.8/  46$ \\ 
\noalign{\vskip 2pt}
a & $  1.23^{+0.00}_{-0.00}$ & $  25.6^{+0.6}_{-0.6}$ & $\ldots$ & $
2.9^{+0.0}_{-0.0}$ & $   4.8^{+ 0.4}_{-  0.3}$ & $\ldots$ & $\ldots$ & $
14.1^{+  2.9}_{- 3.7}$ & $  22 $ & $ 68.1/  39$ \\  
\noalign{\vskip 2pt}
b & $  1.24^{+0.00}_{-0.00}$ & $  28.2^{+  0.6}_{-  0.7}$ & $\ldots$ & $
2.9^{+0.0}_{-0.0}$ & $3.7^{+0.4}_{-0.4}$ & $\ldots$ & $\ldots$ & $1.6^{+
  0.4}_{-  0.4}$ & $ 24 $ & $ 66.6/  39$ \\  
\noalign{\vskip 2pt}
c & $  1.25^{+0.00}_{-0.00}$ & $  29.9^{+0.6}_{-0.6}$ & $\ldots$ &
$3.0^{+0.0}_{-0.0}$ & $ 3.8^{+0.4}_{-0.4}$ & $\ldots$ & $\ldots$ & $
2.0^{+  0.3}_{-  0.3}$ & $  30 $ & $99.6/  39$ \\  
\noalign{\vskip 2pt}
d & $  1.25^{+0.00}_{-0.00}$ & $  29.7^{+  0.6}_{-  0.5}$ & $\ldots$ & $
3.3^{+0.1}_{-0.1}$ & $  5.0^{+0.6}_{-0.6}$ & $\ldots$ & $\ldots$ &  $
1.8^{+  0.4}_{-  0.3}$ & $  28 $ & $ 61.8/  39$ \\  
\noalign{\vskip 2pt}
e & $  1.24^{+0.00}_{-0.00}$ & $  30.5^{+  0.6}_{-  0.5}$ & $\ldots$ & $
3.4^{+0.1}_{-0.1}$ & $4.6^{+ 0.7}_{-  0.6}$ & $\ldots$ & $\ldots$ & $
1.9^{+  0.3}_{-  0.4}$ & $32$ & $ 68.6/  39$ 
\\ 
\noalign{\vskip 2pt}
f & $  1.24^{+0.00}_{-0.00}$ & $  30.9^{+0.6}_{-0.6}$ & $\ldots$ & $
3.4^{+0.1}_{-0.1}$ & $   4.4^{+0.7}_{-  0.6}$ & $\ldots$ & $\ldots$ & $
1.5^{+  0.3}_{- 0.4}$ & $  25 $ & $ 59.3/  39$ \\  
\noalign{\vskip 2pt}
g & $  1.24^{+0.01}_{-0.01}$ & $30.5^{+0.8}_{-0.5}$ & $\ldots$ &
$3.4^{+0.1}_{-0.1}$ & $   4.4^{+ 1.1}_{-  0.9}$ & $\ldots$ & $\ldots$ & $
1.6^{+  0.5}_{-  0.5}$ & $  28 $ & $43.5/  39$ 
\\  
\noalign{\vskip 2pt}
h & $  1.23^{+0.00}_{-0.00}$ & $30.8^{+0.5}_{-0.6}$ & $\ldots$ &
$3.3^{+0.1}_{-0.1}$ & $   4.0^{+0.5}_{-0.5}$ & $\ldots$ & $\ldots$ & $
1.4^{+  0.3}_{-  0.3}$ & $  23 $ & $128/39$ \\  
\noalign{\vskip 2pt}
i & $  1.23^{+0.00}_{-0.00}$ & $  30.8^{+0.7}_{-0.7}$ & $\ldots$ &
$3.1^{+0.1}_{-0.1}$ & $   3.3^{+  0.6}_{-  0.5}$ & $\ldots$ & $\ldots$ & $
1.9^{+  0.3}_{-  0.4}$ & $  30 $ & $54.5/  39$ 
\\ 
\noalign{\vskip 2pt}
j & $1.24^{+0.00}_{-0.00}$ & $29.7^{+0.5}_{-0.5}$ & $\ldots$ &
$3.1^{+0.0}_{-0.1}$ & $4.1^{+0.4}_{-0.4}$ & $\ldots$ & $\ldots$ &
$1.9^{+0.3}_{-0.3}$ & $31$ & $ 63.7/39$ \\   
\noalign{\vskip 2pt}
k & $  1.24^{+0.00}_{-0.00}$ & $  28.8^{+  0.5}_{-  0.6}$ & $\ldots$ & $
3.2^{+ 0.0}_{- 0.1}$ & $4.8^{+0.5}_{-0.5}$ & $\ldots$ & $\ldots$
&$1.4^{+0.3}_{-0.3}$ & $23$ & $123/39$ 
\\  
\noalign{\vskip 2pt}
total $^{\rm b}$ 
& $  1.16^{+ 0.04}_{- 0.02}$ & $\ldots$ & $  2.5^{+ 0.2}_{-  1.1}$ &
$\ldots$ & $\ldots$ & $62^{+ 34}_{-  12}$ & $6.1^{+ 0.1}_{-  0.2}$ &
$10.1^{+ 3.6}_{- 4.6}$ & $ 170 $ & $ 39.5/  46$ \\ 
\noalign{\vskip 2pt}
a & $  1.11^{+0.01}_{-0.01}$ & $\ldots$ & $  5.6^{+0.4}_{-0.1}$ & $\ldots$
  & $\ldots$ & $46^{+5}_{-3}$  & $9.5^{+0.1}_{-0.0}$ & $13.8^{+1.4}_{-1.4}$
  & 230 & $ 57.2/  39$ \\  
\noalign{\vskip 2pt}
b & $1.14^{+0.01}_{-0.01}$ & $\ldots$ & $4.6^{+0.6}_{-0.7}$ & $\ldots$ &
$\ldots$ & $48^{+8}_{-5}$ & $8.3^{+0.2}_{-0.1}$ & $13.5^{+1.7}_{-1.7}$ &
$213$ & $ 46.4/  39$ \\  
\noalign{\vskip 2pt}
c & $1.16^{+0.01}_{-0.01}$ & $\ldots$ & $2.8^{+0.4}_{-0.2}$ & $\ldots$ &
$\ldots$ & $64^{+14}_{-6}$ & $6.9^{+0.2}_{-0.1}$ & $14.6^{+1.5}_{-1.5}$ & $228$
& $ 88.7/  39$ \\  
\noalign{\vskip 2pt}
d & $1.16^{+0.00}_{-0.00}$ & $\ldots$ & $3.3^{+0.1}_{-0.4}$ & $\ldots$ &
$\ldots$ & $43^{+14}_{-5}$ & $5.3^{+0.1}_{-0.2}$ & $13.7^{+1.6}_{-1.5}$ &
$223$ & $ 71.7/  39$ \\  
\noalign{\vskip 2pt}
e & $  1.17^{+0.00}_{-0.01}$ & $\ldots$ & $1.3^{+0.3}_{-0.5}$ & $\ldots$ &
$\ldots$ & $84^{+35}_{-15}$ & $4.5^{+0.1}_{-0.2}$ & $13.1^{+1.5}_{-1.6}$ &
$228$ & $ 80.8/  39$ \\ 
\noalign{\vskip 2pt}
f & $  1.16^{+0.00}_{-0.00}$ & $\ldots$ & $2.9^{+0.1}_{-0.5}$ & $\ldots$ &
$\ldots$ & $40^{+11}_{-5}$ & $4.3^{+0.1}_{-0.1}$ & $11.0^{+1.6}_{-2.2}$ & $190$
& $ 67.9/  39$ \\  
\noalign{\vskip 2pt}
g & $  1.16^{+0.01}_{-0.01}$ & $\ldots$ & $2.1^{+0.3}_{-1.0}$ & $\ldots$ &
$\ldots$ & $50^{+20}_{-10}$ & $4.2^{+0.2}_{-0.2}$ & $10.8^{+3.0}_{-2.2}$ &
$190$ & $ 57.3/  39$ \\ 
\noalign{\vskip 2pt}
h & $  1.15^{+0.01}_{-0.00}$ & $\ldots$ & $2.8^{+0.0}_{-0.4}$ & $\ldots$ &
$\ldots$ & $44^{+11}_{-4}$ & $4.6^{+0.1}_{-0.1}$ & $11.4^{+1.3}_{-1.3}$ & $202$
& $113/  39$ \\  
\noalign{\vskip 2pt}
i & $  1.16^{+0.01}_{-0.01}$ & $\ldots$ & $1.5^{+0.0}_{-0.1}$ & $\ldots$ &
$\ldots$ & $89^{+30}_{-12}$ & $5.6^{+0.2}_{-0.1}$ & $12.2^{+1.7}_{-1.7}$ &
$213$ & $ 68.8/  39$ \\  
\noalign{\vskip 2pt}
j & $  1.16^{+0.00}_{-0.01}$ & $\ldots$ & $2.4^{+0.2}_{-0.1}$ & $\ldots$ &
$\ldots$ & $64^{+15}_{-6}$ & $6.0^{+0.1}_{-0.1}$ & $12.4^{+1.3}_{-1.4}$ & $209$
& $ 80.1/  39$ \\  
\noalign{\vskip 2pt}
k & $1.15^{+0.01}_{-0.00}$ & $\ldots$ & $3.1^{+0.2}_{-0.1}$ & $\ldots$ &
$\ldots$ & $50^{+8}_{-6}$ & $5.8^{+0.1}_{-0.1}$ & $12.2^{+1.3}_{-1.2}$ &
$210$ & $130/  39$ \\  
\noalign{\vskip 2pt}
%
%\hline
\end{tabular}
\end{center}

\medskip
$T_{\rm in}, ~A_{\rm disc}$: Peak multi-temperature disc temperature and
normalization. $\Gamma$: Photon index of the power law. $A_{\rm PL}$: Power
law normalization (photons\,keV$^{-1}$\,cm$^{-2}$\,s$^{-1}$ at 1\,keV).
$A_{\rm compps}$: Normalization of {\tt compps} model. $kT_{\rm e}$:
Coronal electron temperature for the {\tt compps} model. $y_{\rm compps}$:
Compton y-parameter for the {\tt compps} model. $A_{\rm Line}$: Line
normalization (photons\,cm$^{-2}$\,s$^{-1}$ in the line). The Gaussian line
was fixed at 6.4\,keV with a width of $\sigma=0.1$\,keV for the
multi-temperature disc plus black body models, or $\sigma=1.4$\,keV for the
{\tt compps} models. EW: line equivalent width. Uncertainties are at the
90\% confidence level for one interesting parameter ($\Delta \chi^2 =
2.71$). The interstellar equivalent column was fixed at $N_{\rm
  H}=3.2\times 10^{20}\,\rm cm^{-2}$. $^{\rm a}$Parameters for a broad iron
line with $\sigma=1.4^{+0.2}_{-0.2}$\,keV. $^{\rm b}$Parameters for a
broad line with $\sigma =1.4_{-0.1}^{+0.2}$. 

\end{table*}

%%% Local Variables: 
%%% mode: latex
%%% TeX-master: t
%%% End: 

%% file: Table3.tex
\begin{table*}
\caption{Results of Spectral Fitting to the LMC X-1
  Data. \protect{\label{tab:x1fits}}}
\begin{center}
\begin{tabular}{lllllllllcr}
%\hline
%
\noalign{\vskip 2pt}
{$kT_{\rm in}$} & {$A_{\rm disc}$} & {$A_{\rm compps}$}
& {$\Gamma$} & {$A_{\rm PL}$} & $kT_{\rm e}$ & $y_{\rm compps}$ & {$\sigma$} &
 {$A_{\rm Line}$} & {EW} & {$\chi^2/\rm dof$} \\
 {keV} & & {$10^{-2}$} & & {$10^{-1}$} & {keV} & $10^{-2}$ & {keV} &
 {$10^{-5}$} &  {eV} & \\ 
\noalign{\vskip 2pt}
%
%\hline
%
\noalign{\vskip 2pt}
$0.91^{+0.01}_{-0.01}$ & $58.0_{-3.0}^{+3.1}$ & $\ldots$ &
$3.1^{+0.1}_{-0.1}$ & $2.1^{+0.6}_{-0.6}$ & $\ldots$ & $\ldots$ &
\textsl{0.1} & $7^{+4}_{-5}$ & $38$ & 71.8/41 \\ 
\noalign{\vskip 2pt}
$0.88^{+0.01}_{-0.01}$ & $76.3_{-6.0}^{+3.7}$ & $\ldots$ &
$2.9^{+0.1}_{-0.1}$ & $1.5^{+0.4}_{-0.5}$ & $\ldots$ & $\ldots$ &
$0.98^{+0.19}_{-0.15}$ & $32_{-10}^{+10}$ & $195$ & 46.4/40 \\
\noalign{\vskip 2pt}
$0.86_{-0.02}^{+0.02}$ & $\ldots$ & $1.9^{+0.0}_{-0.3}$  & $\ldots$
& $\ldots$ & $68^{+7}_{-8}$ & $4.4^{+0.1}_{-0.3}$ & $0.97^{+0.18}_{-0.14}$
& $33_{-9}^{+10}$ & $207$ & 48.5/40 \\ 
\noalign{\vskip 2pt}
%
%\hline
\end{tabular}
\end{center}

\medskip
See Table.~\ref{tab:x3fits} for an explanation of the symbols. The
interstellar equivalent column was fixed at $N_{\rm H}=7.2\times
10^{21}\,\rm cm^{-2}$. Parameters typeset in italics were frozen at the
indicated value. Errors are at the 90\% confidence level for one
interesting parameter ($\Delta \chi^2 = 2.71$). 

\end{table*}
%%% Local Variables: 
%%% mode: latex
%%% TeX-master: t
%%% End: 

%% file: Table4.tex
\begin{table*}
\caption{\small Results of fitting the \xone\ and \xthree\ 
\asca\ data. \protect{\label{tab:ascafit}}} 
\smallskip
\center{
\begin{tabular}{cccccccc}
\hline
\noalign{\vskip 2pt}
Source & ${\rm N_H}$ & $kT_{\rm e}$ & $A_{\rm compps}$ & 
$y_{\rm compps}$ & $kT_{\rm ray}$ & $A_{\rm ray}$ & $\chi^2$/dof \\
& $\times 10^{22}~ {\rm cm^{-2}}$ & keV & $10^{-2}$ & $10^{-2}$ & keV &
$10^{-2}$ &
\\
\noalign{\vskip 2pt}
\hline
\noalign{\vskip 2pt}
LMC~X-1 & \errtwo{0.63}{0.01}{0.01} & \errtwo{0.75}{0.00}{0.01} &
\errtwo{6.2}{0.0}{0.0} & \errtwo{10.1}{0.1}{0.4} & 
$\ldots$ & $\ldots$ & 498/464 \\
\noalign{\vskip 2pt}
(1995 Apr. 12) & {\it 0.72} & \errtwo{0.67}{0.01}{0.01} &
\errtwo{8.7}{0.2}{0.3} & \errtwo{12.2}{0.3}{0.3} 
& $\ldots$ & $\ldots$ & 924/465 \\ 
\noalign{\vskip 2pt}
& {\it 0.72} & \errtwo{0.71}{0.01}{0.00} &
\errtwo{7.2}{0.2}{0.3} & \errtwo{10.9}{0.4}{0.4} 
& \errtwo{0.82}{0.02}{0.03} &
\errtwo{2.5}{0.2}{0.3} & 523/463 \\ 
\noalign{\vskip 2pt}
LMC~X-3 & \errtwo{0.06}{0.01}{0.00} & \errtwo{0.83}{0.01}{0.00} &
\errtwo{2.7}{0.3}{0.0} & \errtwo{11.2}{0.6}{0.5} 
& $\ldots$ & $\ldots$ & 295/429 \\
\noalign{\vskip 2pt}
(1995 Apr. 15) \\
\noalign{\vskip 2pt}
LMC~X-3 & \errtwo{0.03}{0.00}{0.01} & \errtwo{0.70}{0.01}{0.01} &
\errtwo{2.0}{0.3}{0.2} & \errtwo{11.9}{1.1}{1.0} & 
$\ldots$ & $\ldots$ & 231/414 \\
\noalign{\vskip 2pt}
(1993 Sep. 23) \\
\noalign{\vskip 2pt}
\hline
\end{tabular}
}

\medskip
See Table.~\ref{tab:x3fits} for an explanation of the symbols. $kT_{\rm
  ray}$, $A_{\rm ray}$ are the temperature and normalization of the
Raymond-Smith plasma component \cite{raymond:77a}. Parameters typeset in
italics were frozen at the indicated value. Errors are at the 90\%
confidence level for one interesting parameter ($\Delta \chi^2 = 2.71$). 

\end{table*}

%%% Local Variables: 
%%% mode: latex
%%% TeX-master: t
%%% End: 

%% file: lmc_long.bbl
\begin{thebibliography}{}

\bibitem[\protect\citename{Arnaud }1996]{arnaud:96a}
Arnaud K.A.,  1996,
\newblock in Jacoby J.H., Barnes J., eds, Astronomical Data Analysis Software
  and Systems {V}. Astron.\ Soc.\ Pacific, Conf.\ Ser. 101 Astron.\ Soc.\
  Pacific, San Francisco, p.~17

\bibitem[\protect\citename{{Ba\l{}uci\'{n}ska}-Church \& {McCammon}
  }1992]{balu:92a}
{Ba\l{}uci\'{n}ska}-Church M., {McCammon} D.,  1992, ApJ, 400, 699

\bibitem[\protect\citename{Bendat \& Piersol }1986]{bendat}
Bendat J., Piersol A.,  1986,
\newblock Random Data: Analysis and Measurement Procedures,
\newblock Wiley, New York

\bibitem[\protect\citename{Benensohn et~al. }1997]{benensohn:97a}
Benensohn J.S., Lamb D.Q., Taam R.E.,  1997, ApJ, 478, 723

\bibitem[\protect\citename{Bondi \& Hoyle }1944]{bondi:44a}
Bondi H., Hoyle F.,  1944, MNRAS, 104, 273

\bibitem[\protect\citename{Bowyer et~al. }1965]{bowyer:65a}
Bowyer S., Byram E.T., Chubb T.A., Friedman H.,  1965, Science, 147, 394

\bibitem[\protect\citename{Brandt et~al. }1996]{brandt:96a}
Brandt W.N., Fabian A.C., Dotani T., et~al., 1996, MNRAS, 283, 1071

\bibitem[\protect\citename{Cowley et~al. }1978]{cowley:78a}
Cowley A.P., Crampton D., Hutchings J.B.,  1978, AJ, 83, 1619

\bibitem[\protect\citename{Cowley et~al. }1983]{cowley:83a}
Cowley A.P., Crampton D., Hutchings J.B., et~al., 1983, ApJ, 272, 118

\bibitem[\protect\citename{Cowley et~al. }1995]{cowley:95a}
Cowley A.P., Schmidtke P.C., Anderson A.L., {McGrath} T.K.,  1995, PASP, 107,
  145

\bibitem[\protect\citename{Cowley et~al. }1991]{cowley:91a}
Cowley A.P., Schmidtke P.C., Ebisawa K., et~al., 1991, ApJ, 381, 526

\bibitem[\protect\citename{Cowley et~al. }1994]{cowley:94a}
Cowley A.P., Schmidtke P.C., Hutchings J.B., Crampton D.,  1994, ApJ, 429, 826

\bibitem[\protect\citename{Cui et~al. }1997]{cui:97b}
Cui W., Zhang S.N., Focke W., Swank J.H.,  1997, ApJ, 484, 383

\bibitem[\protect\citename{Day et~al. }1998]{day:98a}
Day C., Arnaud K., Ebisawa K., et~al., 1998,
\newblock The {ASCA} Data Reduction Guide,
\newblock Technical report, NASA Goddard Space Flight Center, Greenbelt, Md.
  Version~2.0

\bibitem[\protect\citename{Dove et~al. }1998]{dove:98a}
Dove J.B., Wilms J., Nowak M.A., et~al., 1998, MNRAS, 298, 729 (paper I)

\bibitem[\protect\citename{Ebisawa et~al. }1993]{ebisawa:93a}
Ebisawa K., Makino F., Mitsuda K., et~al., 1993, ApJ, 403, 684

\bibitem[\protect\citename{Ebisawa et~al. }1989]{ebisawa:89a}
Ebisawa K., Mitsuda K., Inoue H.,  1989, PASJ, 41, 519

\bibitem[\protect\citename{Fryxell \& Taam }1988]{fryxell:88a}
Fryxell B.A., Taam R.E.,  1988, ApJ, 335, 862

\bibitem[\protect\citename{{Gierli\'nski} et~al. }1999]{gierlinski:99a}
{Gierli\'nski} M., Zdziarski A.A., Poutanen J., et~al., 1999, MNRAS,
309, 496 

\bibitem[\protect\citename{Grebenev et~al. }1993]{grebenev:93a}
Grebenev S., Sunyaev R., Pavlinsky M., et~al., 1993, A\&AS, 97, 281

\bibitem[\protect\citename{Hoyle \& Lyttleton }1939]{hoyle:39a}
Hoyle F., Lyttleton R.A.,  1939,
\newblock Proc. Cam. Phil Soc., 35, 405

\bibitem[\protect\citename{Hutchings et~al. }1983]{hutchings:83a}
Hutchings J.B., Crampton D., Cowley A.P.,  1983, ApJ, 275, L43

\bibitem[\protect\citename{Hutchings et~al. }1987]{hutchings:87a}
Hutchings J.B., Crampton D., Cowley A.P., et~al., 1987, AJ, 94, 340

\bibitem[\protect\citename{Jahoda et~al. }1996]{jahoda:96b}
Jahoda K., Swank J.H., Giles A.B., et~al., 1996,
\newblock in Siegmund O.H., ed, {EUV}, X-Ray, and Gamma-Ray Instrumentation for
  Astronomy {VII}. Proc.\ SPIE 2808 SPIE, Bellingham, WA, p.59

\bibitem[\protect\citename{Lampton et~al. }1976]{lampton:76a}
Lampton M., Margon B., Bowyer S.,  1976, ApJ, 208, 177

\bibitem[\protect\citename{Leahy et~al. }1983]{leahy:83a}
Leahy D.A., Darbro W., Elsner R.F., et~al., 1983, ApJ, 266, 160

\bibitem[\protect\citename{Long et~al. }1981]{long:81a}
Long K.S., Helfand D.J., Grabelsky D.A.,  1981, ApJ, 248, 925

\bibitem[\protect\citename{Merloni et~al. }1999]{merloni:99a}
Merloni A., Fabian A.C., Ross R.R.,  1999, MNRAS, 313, 193

\bibitem[\protect\citename{Mitsuda et~al. }1984]{mitsuda:84a}
Mitsuda K., Inoue H., Koyama K., et~al., 1984, PASJ, 36, 741

\bibitem[\protect\citename{Miyamoto et~al. }1994]{miyamoto:94a}
Miyamoto S., Kitamoto S., Iga S., et~al., 1994, ApJ, 435, 398

\bibitem[\protect\citename{{Nagase} et~al. }1994]{nagase:94a}
{Nagase} F., {Zylstra} G., {Sonobe} T., et~al., 1994, ApJ, 436, L1

\bibitem[\protect\citename{Nowak }1995]{nowak:95a}
Nowak M.A.,  1995, PASP, 107, 1207

\bibitem[\protect\citename{Nowak et~al. }1999a]{nowak:99a}
Nowak M.A., Vaughan B.A., Wilms J., et~al., 1999a, ApJ, 510, 874

\bibitem[\protect\citename{Nowak et~al. }1999b]{nowak:99c}
Nowak M.A., Wilms J., Dove J.B.,  1999b, ApJ, 517, 355

\bibitem[\protect\citename{Nowak \& Wilms }1999]{nowak:99d}
Nowak M.A., Wilms J.,  1999, ApJ, 522, 476

\bibitem[\protect\citename{Nowak et~al. }1999c]{nowak:99b}
Nowak M.A., Wilms J., Vaughan B.A., et~al., 1999c, ApJ, 515, 726

\bibitem[\protect\citename{Poutanen \& Svensson }1996]{poutanen:96a}
Poutanen J. \&  Svensson R., 1986, ApJ, 470, 249

\bibitem[\protect\citename{Raymond \& Smith}{1977}]{raymond:77a}
Raymond, J.~C., \& Smith, B.~W.,  1977, ApJS, 35, 419

\bibitem[\protect\citename{Rothschild et~al. }1998]{rothschild:98a}
Rothschild R.E., Blanco P.R., Gruber D.E., et~al., 1998, ApJ, 496, 538

\bibitem[\protect\citename{Ruffert }1997]{ruffert:97a}
Ruffert M.,  1997, AJ, 317, 793

\bibitem[\protect\citename{Ruffert }1999]{ruffert:99a}
Ruffert M.,  1999, AJ, 346, 861

\bibitem[\protect\citename{Schandl }1996]{schandl:96a}
Schandl S.,  1996, A\&A, 307, 95

\bibitem[\protect\citename{Schlegel et~al. }1994]{schlegel:94a}
Schlegel E.M., Marshall F.E., Mushotzky R.F., et~al., 1994, ApJ, 422, 243

\bibitem[\protect\citename{Schmidtke et~al. }1999]{schmidtke:99a}
Schmidtke P.C., Ponder A.L., Cowley A.C.,  1999, AJ, 117, 1292

\bibitem[\protect\citename{Shields et~al. }1986]{shields:86a}
Shields G.A., McKee C.F., Lin D.N.C.,  1986, ApJ, 306, 90

\bibitem[\protect\citename{Syunyaev et~al. }1990]{sunyaev:90a}
Syunyaev R.A., {Gil'fanov} M., Churazov E., et~al., 1990, Sov. Astron. Lett.,
  16, 55

\bibitem[\protect\citename{Taam \& Fryxell }1988]{taam:88a}
Taam R.E., Fryxell B.,  1988, ApJ, 327, L73

\bibitem[\protect\citename{Taam \& Fryxell }1989]{taam:89a}
Taam R.E., Fryxell B.,  1989, ApJ, 339, 297

\bibitem[\protect\citename{Tanaka \& Lewin }1995]{tanaka:95a}
Tanaka Y., Lewin W.H.G.,  1995,
\newblock Black-hole binaries. in Lewin W.H.G., {van Paradijs} J., {van den
  Heuvel} E.P.J., eds, {X}-Ray Binaries. Cambridge Univ.\ Press, Cambridge,
  Ch.~3, p. 126

\bibitem[\protect\citename{Treves et~al. }1988]{treves:88a}
Treves A., Belloni T., Chiapetti L., et~al., 1988, ApJ, 325, 119

\bibitem[\protect\citename{Treves et~al. }1990]{treves:90a}
Treves A., Belloni T., Corbet R.H.D., et~al., 1990, ApJ, 364, 266

\bibitem[\protect\citename{{van der Klis} }1989]{vanderklis:89b}
{van der Klis} M.,  1989,
\newblock in {\"Ogelman} H., {van den Heuvel} E.P.J., eds, Timing Neutron
  Stars. NATO ASI C262 Kluwer, Dordrecht, p.27

\bibitem[\protect\citename{{van der Klis} et~al. }1983]{vanderklis:83a}
{van der Klis} M., Tjemkes S., {van Paradijs} J.,  1983, A\&A, 126, 265

\bibitem[\protect\citename{Vaughan \& Nowak }1997]{vaughan:97a}
Vaughan B.A., Nowak M.A.,  1997, ApJ, 474, L43

\bibitem[\protect\citename{Wilms et~al. }1999a]{wilms:99aa}
Wilms J., Nowak M.A., Dove J.B., et~al., 1999a, ApJ, 522, 460

\bibitem[\protect\citename{Wilms et~al. }1999b]{wilms:99a}
Wilms J., Nowak M.A., Dove J.B., et~al., 1999b,
\newblock in Aschenbach B., Freyberg M., eds, {H}ighlights in {X}-ray
  {A}stronomy. MPE Report 272, p.~98

\bibitem[\protect\citename{Wilms et~al. }1999c]{wilms:99c}
Wilms J., Nowak M.A., Dove J.B., et~al., 1999c, Astrophys.\ Lett.\
Comm., 38, 273

\bibitem[\protect\citename{Wilms et~al. }1999d]{wilms:99b}
Wilms J., Nowak M.A., Heindl W.A., et~al., 1999d, MNRAS submitted (Paper II)

\bibitem[\protect\citename{Zhang et~al. }1997]{zhang:97b}
Zhang S.N., Cui W., Harmon B.A., et~al., 1997, ApJ, 477, L95

\bibitem[\protect\citename{Zhang \& Jahoda }1996]{zhangw:96a}
Zhang W., Jahoda K.,  1996,
\newblock Deadtime Effects in the {PCA},
\newblock Technical report, NASA GSFC, Greenbelt

\bibitem[\protect\citename{Zhang et~al. }1995]{zhangw:95a}
Zhang W., Jahoda K., Swank J.H., et~al., 1995, ApJ, 449, 930

\end{thebibliography}
